\documentclass[aps,twocolumn,prc,superscriptaddress,noshowpacs,nofootinbib,noshowkeys,floatfix]{revtex4-1}
\usepackage[colorlinks=true,linktocpage=true,linkcolor=blue,citecolor=blue]{hyperref}
\usepackage[usenames,dvipsnames]{color}
\usepackage{amsmath, amssymb}
\usepackage{multirow}
\usepackage{longtable}
\usepackage{comment}
\usepackage{color}
\usepackage{xcolor}
\usepackage{xspace}
\usepackage{cleveref}
\usepackage[normalem]{ulem}  
\usepackage{graphicx,lipsum}
\usepackage{ragged2e}
\usepackage{array} 

\newcommand{\sqsn}{\mbox{$\sqrt{s_{_{\rm{NN}}}}$}\xspace}
\newcommand{\pt}{$p_{\rm{T}}$\xspace}

\newcommand{\hfs}{heavy-flavors\xspace}
\newcommand{\pta}{$p_{\rm{T}}^{\rm assoc}$\xspace}
\newcommand{\ptt}{$p_{\rm{T}}^{\rm trig}$\xspace}

\newcommand{\ptaB}{$p_{\rm{T}}^{\rm \bar{B}}$\xspace}
\newcommand{\pttB}{$p_{\rm{T}}^{\rm B}$\xspace}
\newcommand{\ptaD}{$p_{\rm{T}}^{\rm \bar{D^{0}}}$\xspace}
\newcommand{\pttD}{$p_{\rm{T}}^{\rm D^{0}}$\xspace}

\newcommand{\gev}{GeV/$c$\xspace}
\newcommand{\delphi}{$\rm{\Delta \varphi}$\xspace}

\newcommand{\bef}{\begin{figure}}
\newcommand{\eef}{\end{figure}}
\newcommand{\bc}{\begin{center}}
\newcommand{\ec}{\end{center}}

\newcommand{\be}{\begin{equation}}
\newcommand{\ee}{\end{equation}}
\newcommand{\bea}{\begin{eqnarray}}
\newcommand{\eea}{\end{eqnarray}}

\newcommand{\dz}{$\rm{D^{0}}$\xspace}

\newcommand{\ddbar}{$\rm{D^0-\Bar{D}^0}$\xspace}

\newcommand{\bbbar}{$\rm{B-\Bar{B}}$\xspace}
\newcommand{\ptaa}{$\rm{2-4 \, GeV/c}$\xspace}

\begin{document}

\title{{\Large An Impact of Parton Distribution Functions on Open Heavy Flavor Azimuthal Angular Correlations in Pb-Pb Collisions at $\sqrt{s_{NN}}$ = 5.5 TeV}}

\author{Swapnesh Khade\footnote{Corresponding author.}}
\email{swapneshkhade@gmail.com}
\affiliation{Department of Physics, Indian Institute of Technology Indore, Simrol, Indore 453552, India}
\author{Ravindra Singh}
\email{ravirathore.physics@gmail.com}
\affiliation{Istituto Nazionale di Fisica Nucleare Sezione di Padova INFN, Padova, Italy}
\author{Yoshini Bailung}
\email{yoshini.bailung.1@gmail.com}
\affiliation{Department of Physics, Indian Institute of Technology Indore, Simrol, Indore 453552, India}
\author{Ankhi Roy}
\email{ankhi@iiti.ac.in}
\affiliation{Department of Physics, Indian Institute of Technology Indore, Simrol, Indore 453552, India}

\begin{abstract}

In this paper, we investigate the impact of modern free parton distribution functions and nuclear parton distribution functions on the azimuthal angular correlation of open heavy flavor hadrons in Pb-Pb collisions at \sqsn = 5.5 TeV using PYTHIA8 + Angantyr model. The method involves a transverse momentum (\pt) differential assessment of azimuthal angular correlations between prompt \ddbar and \bbbar. By analyzing the \pt-dependent behavior of these correlations, we probe the interplay between different production mechanisms and assess the impact of parton distribution functions on heavy-flavor hadron production dynamics.
Our results indicate that nuclear parton distribution functions significantly modify both near-side and away-side peaks of the azimuthal correlation distribution compared to the default PYTHIA8 baseline, particularly at low \pt. The near-side double-peak structure suggests contributions from next-to-leading order (NLO) processes, such as gluon splitting, and its presence is strongly correlated with the number of hard multi-partonic interactions. Additionally, by quantitatively comparing the change in correlation widths across various PDF settings, we comment on the role of nuclear parton distribution functions in heavy-quark thermalization in heavy-ion collisions. This comprehensive study of azimuthal correlations enhances our understanding of initial-state conditions and heavy-flavor production dynamics in nuclear collisions.

\end{abstract}
\date{\today}
\maketitle

\section{Introduction}
\label{intro}

Experiments at the Large Hadron Collider at CERN, Geneva, and the the Relativistic Heavy Ion Collider (RHIC) at BNL, USA, explore the properties of quark-gluon plasma (QGP), an exotic state of matter, by colliding ultra-relativistic heavy-ions~\cite{STAR:2005gfr,Braun-Munzinger:2015hba,ALICE:2022wpn,ALICE:2016fzo}. The QGP was discovered at the SPS, in Pb-Pb collisions at center of mass energy \sqsn = 17.3 GeV~\cite{NA50:2000brc}, inspiring future experiments such as ALICE and STAR to perform precise measurements and understand its properties.

An evolution of heavy-ion collision comprises an initial or pre-equilibrium state, a deconfined state of partons, followed by hadronization, and ultimately, freeze-out of the final state hadrons~\cite{ALICE:2022wpn}. To systematically study all the stages of a heavy-ion collision, it is necessary to use a probe that is produced at the onset of the collision and traverses through the various stages. Heavy quarks (charm and beauty), due to their large masses, are produced at the early stages of heavy-ion collision via initial hard scatterings. They experience a complete history of the heavy-ion collision and serve as excellent probes to study its various stages. While interacting with the pre-equilibrium state and the QGP medium, heavy quarks lose energy via collisional and radiative processes, which are followed by fragmentation~\cite{Bonino:2023icn,Kartvelishvili:1977pi, Andersson:1983jt, ALICE:2021dhb} and hadronization processes~\cite{Zhao:2023nrz, Fries:2008hs,Frixione:1997ma, Norrbin:2000zc, Braun-Munzinger:2003pwq, Alberico:2013bza}. In hadronic collisions, the production of open heavy flavor hadrons~\cite{Fujii:2013yja} can be explained using the factorization theorem. For open charm hadrons, the total production cross-section is given by,

\begin{align}
d\sigma_{AB\rightarrow C}^{\rm{hard}} &= \Sigma_{a,b,X} f_{a/A} (x_a,Q^2) \otimes f_{b/B} (x_b,Q^2) \notag \\
&\quad \otimes d\sigma_{ab\rightarrow cX}^{\rm{hard}} (x_a,x_b,Q^2) \otimes D_{c\rightarrow C}(z,Q^2)
\label{eq:factor}
\end{align}

where, $f_{a/A} (x_a, Q^2)$ and $f_{b/B} (x_b, Q^2)$ are the parton distribution functions (PDFs) which give the probability of finding parton $a(b)$ inside the hadron $A(B)$ for a given $x$, where $x$ is the fraction of proton momentum taken by a parton, and $Q^2$ is the factorization scale. $d\sigma_{ab\rightarrow cX}^{\rm{hard}} (x_a,x_b, Q^2)$ is the partonic hard scattering cross-section, and $D_{c\rightarrow C}(z, Q^2)$ is the fragmentation function of the produced hadron. For beauty quarks, the production of B-mesons will have effects due to its higher mass than charm, which alters production cross-section and its fragmentation function~\cite{Kniehl:2007erq, Salajegheh:2019ach, Epele:2018ewr}. A partonic hard scattering cross-section is calculated by perturbative quantum chromodynamic (pQCD) calculations, while the fragmentation function and PDFs invoke non-perturbative treatments. The production of heavy quarks has contributions from gluon fusion and pair annihilation, which are leading order (LO) processes, while gluon splitting and flavor excitation correspond to processes from the next-to-leading order (NLO). At \sqsn below $\rm{10}$ GeV, the pair creation and flavor excitation processes dominate the heavy quark production, while a negligible contribution is expected from gluon splitting.  At the LHC and the highest RHIC energies, the NLO processes dominate charm quark production over the LO processes. For beauty quarks, flavor excitation has the highest contribution, while gluon splitting contributes the least among the three production processes. This change in the contribution from different NLO production processes can be attributed to an increase in available phase space in the initial and final state showers as a function of the center of mass energy. The contrast of these contributions show sensitivity to the quark transverse momentum (\pt) as well its pseudo-rapidity ($\rm{\eta}$)~\cite{Norrbin:2000zc,Hwa:1979pn,Biro:1994mp}.

Parton distribution functions are extracted using a global analysis of data obtained from ep, pp, and p-Pb collision systems. Free proton PDFs and nuclear PDFs (nPDFs) are the two types of PDFs that are possible to obtain from data. Nuclear PDFs are modified free proton PDFs due to the shadowing, anti-shadowing, and EMC effect ~\cite{EuropeanMuon:1992pyr} in the presence of a nuclear environment. Many free proton PDFs and nPDFs extracted by various groups are available in the LHAPDF6~\cite{Buckley:2014ana} database. ATLAS measured the nuclear modification factor of heavy flavors in p-Pb collisions at \sqsn = 5.02 TeV, and 8.16 TeV, at forward rapidities~\cite{LHCb:2017yua,LHCb:2022dmh}, revealing the impact of nuclear shadowing at low \pt. These results are consistent with predictions made using nPDFs~\cite{Kusina:2017gkz, ALICE:2020mso}. A comparison of rapidity(y) integrated $\mathrm{W^{\pm}/Z}$-boson production by ATLAS~\cite{ATLAS:2019maq} and ALICE~\cite{ALICE:2020jff} with theoretical predictions using nPDFs show a 3.4$\sigma$ deviation from free proton PDFs~\cite{ALICE:2022wpn}. This emphasizes the significance of nuclear PDFs in heavy-ion collisions, particularly at forward rapidities. We anticipate that heavy quark production will be particularly sensitive to the employment of nPDFs in heavy-ion collisions due to the lower factorization scales involved in production compared to $\mathrm{W^{\pm}/Z}$ bosons.

In heavy-ion collisions, measurements of heavy flavor nuclear modification factor ($\mathrm{R_{AA}}$)~\cite{CMS:2017qjw,ALICE:2015zhm,ALICE:2019hno,ALICE:2015vxz,ALICE:2018lyv,STAR:2014wif} and elliptic flow ($\mathrm{v_2}$) of prompt and non-prompt charm hadrons~\cite{ALICE:2023gjj,STAR:2017kkh,ALICE:2014qvj,ALICE:2013olq,ALICE:2020hdw} have provided ample information about energy loss mechanisms and thermalization of heavy quarks in the QGP. The measurement of charm $\mathrm{v_{2}}$ as a function of \pt by ALICE~\cite{ALICE:2013olq,ALICE:2020hdw,ALICE:2023gjj}, quotes their participation in the collective medium, corresponding to its thermalization in the region $2 < p_{\mathrm{T}}< 8 $ \gev. Although this measurement confirms the thermalization of charm quarks in the QGP, it cannot comment on the degree of thermalization, viz. fraction of thermalized charm quarks. The ALICE3 experiment investigates the thermalization of charm quarks by measuring the azimuthal angular correlation of $\rm{D^0-\bar{D}^0}$~\cite{ALICE:2022wwr}. The azimuthal angular correlation of \ddbar can be an excellent observable in this regard, which will allow one to look for effects of thermalization in differential \pt ranges. The azimuthal angular correlation~\cite{Beraudo:2014boa, PHENIX:2018wex, Zhang:2019bkf, Zhang:2018ucx,ALICE:2019oyn,ALICE:2021kpy,ALICE:2016clc}, is the distribution of the differences in azimuthal angles, $\rm{\Delta \varphi = \varphi_{trig}-\varphi_{assoc}}$, and pseudorapidities, $\rm{\Delta \eta = \eta_{trig} - \eta_{assoc}}$, where $\rm{\varphi_{trig}}$ ($\rm{\eta_{trig}}$) and $\rm{\varphi_{assoc}}$ ($\rm{\eta_{assoc}}$) are the azimuthal angles (pseudorapidities) of the trigger \footnote[1]{~reference particle with respect to which correlation is computed.} and associated particles \footnote[2]{~ particles being correlated with the trigger.}, respectively. The structure of the correlation function usually contains a $``$near-side'' (NS) peak and an $``$away-side'' (AS) peak at $\rm{\Delta \varphi=0}$ and $\rm{\Delta \varphi=\pi}$ respectively over a wide range of $\rm{\Delta \eta}$. For fully thermalized charm quarks, the flight direction will be completely randomized, leading to flat \ddbar azimuthal angular correlation~\cite{ALICE:2022wwr}, and any deviation from flatness can be used to estimate the degree of thermalization. Besides thermalization, \ddbar and \bbbar azimuthal angular correlation is an excellent probe of heavy flavor production mechanisms. The LO processes are important for the AS peak, while the NLO process like gluon splitting is a major contributor to the NS of \ddbar correlation. A \pt dependent study will allow disentangling the contribution of pair creation and gluon splitting processes in charm production and comment on the degree of thermalization. 

In this contribution, we study the effects of nPDFs on the azimuthal angular correlation of prompt \ddbar, and \bbbar in Pb-Pb collisions at $\sqrt{s_{NN}} = 5.5$ TeV. We use the PYTHIA8+Angantyr model~\cite{Sjostrand:2006za,Bierlich:2018xfw} to study non-collective effects and provide an effective comment on heavy quark thermalization in heavy ion collisions. Because PYTHIA8 lacks a thermal medium, it is an ideal platform to isolate and study the effects of nPDFs, and make conclusive remarks about the thermalization of heavy flavors. Since nPDFs can affect the production of heavy quarks during initial and final state radiations (ISR and FSR) via hard scatterings, their imprints can reflect in the azimuthal angular correlation of open charm and beauty mesons. The comparison of yields obtained from azimuthal angular correlation of \dz and $\rm{B}$ mesons can give us flavor-dependent and \pt-dependent relative contributions of gluon splitting and pair creation processes.

This article is organized as follows, In Section~\ref{section2}, we briefly discuss about PDFs. Section~\ref{section3} contains a detailed discussion of the event generation, methodology, and analysis strategy. Section ~\ref{section4} includes the results of the study. Important findings of the study are summarized in section~\ref{section5}.

\section{Parton distribution functions}
\label{section2}

Parton distribution functions play a pivotal role in hadron and particle physics, as they are crucial for interpreting experimental data from various high-energy processes. PDFs cannot be derived from first principles; instead, they are determined by comparing theoretical predictions of hadronic cross-sections with experimental data. These functions are modeled using fit parameters optimized through data to achieve matching theoretical predictions. At an input interaction scale \( Q^2_0 \) of partons, PDFs are parameterized for each parton via the equation:
\begin{align}
f^p(x,Q^2_0)=\mathcal{N}x^{\alpha_{f}}(1-x)^{\beta_f}\mathcal{I}(x;a)
\label{eq.parameteriation}
\end{align}
where \( x \) is the Bjorken variable, \( \alpha_{f} \) and \( \beta_f \) ~\cite{Ball:2016spl} are parameters obtained from data in global QCD analyses. The factor \( \mathcal{N} \) is a normalization constant that accounts for theoretical constraints. The function \( \mathcal{I}(x; a) \), depending on the set parameters $``a$'', is designed to interpolate between small and large \( x \)-values. The power-like factors \( x^{\alpha_f} \) and \( (1-x)^{\beta_f} \) describe the low-\( x \) and high-\( x \) behavior of PDFs, inspired by Regge theory ~\cite{Regge:1959mz} and the Brodsky-Farrar quark counting rules~\cite{Brodsky:1973kr} , respectively.

As previously mentioned, the parameterization of PDFs is set at an input scale \( Q_0 \); these PDFs subsequently evolve to the scale \( Q \) using the DGLAP equation~\cite{Gribov:1972ri,Lipatov:1974qm,Altarelli:1977zs,Dokshitzer:1977sg}. Predictions for hadronic cross-sections are obtained by convoluting the evolved PDFs with partonic cross-sections at a given perturbative order. Optimal PDF parameters are found using a suitable optimization method, commonly the log-likelihood or \(\chi^2\) minimization technique.

Nuclear PDFs are defined as the average of proton and neutron densities in a nucleus~\cite{Barshay:1975zz,Ethier:2020way}:
\begin{align}
    f^{(A,Z)}_{i}(x,Q^2)=\frac{Z}{A}f^{p/A}_{i}(x,Q^2)+\frac{(A-Z)}{A}f^{n/A}_{i}(x,Q^2)
    \label{eq.nuclear_pdf1}
\end{align}
In this expression, the pair of atomic numbers $(A, Z)$ identifies the nuclear isotopes, and $f^{p,n/A}_{i}$ are the proton and neutron-bound PDFs, related to the nucleon PDFs by
\begin{align}
    f^{p,n/A}_{i}(x,Q^2)=R^A_{i}(x,Q^2)f^{p,n}_{i}(x,Q^2)
    \label{eq.nuclear_pdf2}
\end{align}

where \( R^A_{i}(x,Q^2) \) is a scale-dependent nuclear modification. This factor illuminates the various nuclear modifications concerning the free proton in different \( x \) regions. Experimental data have indicated a shadowing behavior (\( R<1 \)) in the low \( x < 0.1 \) region, followed by anti-shadowing (\( R>1 \)) around \( x \approx 0.1 \), and the EMC effect (\( R<1 \)) in the valence region, \( x \approx 0.4 \)~\cite{EuropeanMuon:1992pyr}. The mechanisms generating these effects are still under investigation but appear as a general feature in all nuclear PDFs.

The free proton PDFs under our consideration are NNPDF2.3~\cite{Ball:2012cx}, NNPDF4.0~\cite{NNPDF:2021njg} and CT18ANLO~\cite{Hou:2019efy} PDF. The NNPDF2.3 is a default proton PDF for which different PYTHIA8 processes are tuned, providing a standard PYTHIA8 baseline for comparison. It is the first global PDF set to systematically include all relevant LHC data up to 2012, employing a variable flavor number scheme based on FONLL ~\cite{Cacciari:1998it} for heavy quarks. The CT18 QCD global fit uses high-statistics LHC datasets, HERA I + II DIS data, and datasets from the CT14 global QCD analysis~\cite{Dulat:2015mca}, in total 40 datasets. The global fit also incorporates ATLAS at \(\sqrt{s}=7\) TeV $W$ and $Z$ boson data. For CT18ANLO, parton PDFs are parameterized using the SACOT-\(\chi\) scheme~\cite{Aivazis:1993pi}.

The nuclear PDFs used are EPPS21~\cite{Eskola:2021nhw} and nNNPDF3~\cite{AbdulKhalek:2022fyi}. The EPPS21 nuclear PDF is derived using CT18ANLO as the free proton PDF, with nuclear modifications parameterized at the charm pole mass threshold \( Q_0 = m_{\mathrm{charm}} = 1.3 \) GeV. This parameterization addresses the EMC effect, anti-shadowing, and shadowing at appropriate \( x \) values. The number of independent PDFs are six, and heavy quarks are treated using the SACOT-\(m_{T}\) general-mass variable flavor number scheme (GM-VFNS). The nNNPDF3 uses Machine Learning to extract PDFs from the data, based on NNPDF4 as the free proton PDF, with heavy quarks treated using the FONLL scheme. The use of nNNPDF3.0 is important for examining the effects of different heavy quark treatments. Both nuclear PDFs use a lot of new data to constrain gluon PDFs compared to their predecessors. The most relevant data related to heavy quark production is the measurement of the nuclear modification factor of \dz mesons in p-Pb collisions by LHCb~\cite{LHCb:2017yua,LHCb:2022dmh}. Since the production of charm quark is a sensitive probe of gluon PDFs, it allows to put a stringent constraint on the gluon PDFs up to $\rm{x \approx 10^{-5}}$.

The above-mentioned PDFs were selected for study due to differences in the theoretical treatment of heavy quarks, the number of fit parameters, the methods used for parameterization, and various other factors.


\section{Event generation and Analysis methodology}
\label{section3}
\subsection{The Angantyr model}

The PYTHIA8 event generator~\cite{Sjostrand:2006za, Singh:2021edu, Sjostrand:2007gs, Bierlich:2018xfw, Skands:2014pea, Campbell:2022qmc, Corke:2010yf, Lonnblad:2021fyl} is used to investigate pp, ep, p-A, and A-A collisions in depth. It uses 2 → 2 QCD matrix elements calculated with leading-order accuracy, incorporating next-to-leading order effects during the parton showering stage. The parton showering follows a leading-logarithmic \pt ordering, with soft-gluon emission divergences prevented by an additional veto. The hadronization process is managed using the Lund string-fragmentation model. It offers a plethora of processes and tunes to use based on the physics involved in the study. PYTHIA8 employs an intial hard scattering with the incoming proton beams, followed by ISR, multi-parton interactions (MPI)~\cite{Corke:2009tk,Diehl:2011yj,Sjostrand:2017cdm} and FSR processes, comprising the underlying event. The high \pt partons give rise to showers or jets that fragment and hadronize according to the Lund string fragmentation model~\cite{Andersson:1983ia}. Hadronization is accomplished through the Color Reconnection (CR) mechanism between fragmenting partons~\cite{Khoze:1994fu,Gieseke:2012ft,Rathsman:1998tp,Bierlich:2015rha}, which is accomplished by rearranging the strings between them. This modifies the total string length, affecting hadronization. When the string length is small enough after the subsequent creation of quark-antiquark combinations, the partons hadronize into hadrons. The MPI and CR phenomena in PYTHIA8 play an essential role in the particle production, as evidenced by the charged-particle multiplicity and \pt distributions, and particle ratios for pp collisions at LHC energies~\cite{Christiansen:2015yqa}. 

Leading order perturbative scattering processes of gluon fusion ($gg \rightarrow Q\overline{Q}$) or pair annihilation ($q\overline{q} \rightarrow Q\overline{Q}$) is used for the production of \hfs in PYTHIA8. PYTHIA8 also approximates certain higher-order contributions within its LO framework via flavor excitations ($gQ \rightarrow Qg$), or gluon splittings ($g\rightarrow Q\overline{Q}$) which give rise to heavy-flavor production during high \pt parton showers~\cite{Braun-Munzinger:2003pwq,Norrbin:2000zc}.

In PYTHIA8, heavy-ion collisions are simulated using the Angantyr model~\cite{Bierlich:2018xfw}, which extrapolates proton-proton (pp) collision dynamics to nucleus-nucleus (AA) collisions. The positions of nucleons within the nucleus are determined by generating a nuclear density distribution using Monte Carlo methods. The number of interacting nucleons and binary collisions is calculated using the Glauber formalism, which is based on the eikonal approximation in impact parameter space. Here, the projectile nucleons are assumed to travel along straight lines, undergoing multiple sub-collisions with nucleons in the target. The contribution from individual nucleon-nucleon interactions in the final state is estimated using the Fritiof model, which accounts for both diffractively and non-diffractively wounded nucleons and manages soft particle production~\cite{Bierlich:2016smv}. Hard particles are produced in individual nucleon-nucleon sub-collisions. At high energies, hard partonic sub-collisions play a crucial role, which is addressed by introducing the concepts of primary and secondary absorptive interactions. The corresponding nucleon-nucleon parton-level events are generated using the full multi-parton interaction (MPI) machinery in PYTHIA8 for both absorptive and diffractive interactions. 

While the Angantyr model does not employ a thermalized medium, it accurately describes the multiplicity of heavy-ion collisions as a function of centrality and rapidity (y) distribution. As a result, Angantyr offers the unique opportunity to categorize the collective and non-collective effects in heavy-ion collisions. As \ddbar correlation measurements are sensitive to heavy quark thermalization in the QGP, our study will allow separating the non-collective components of heavy-flavor production and dynamics in heavy-ion collisions.
\begin{figure}[h!]
  \raggedright
  \includegraphics[width=0.45\textwidth]{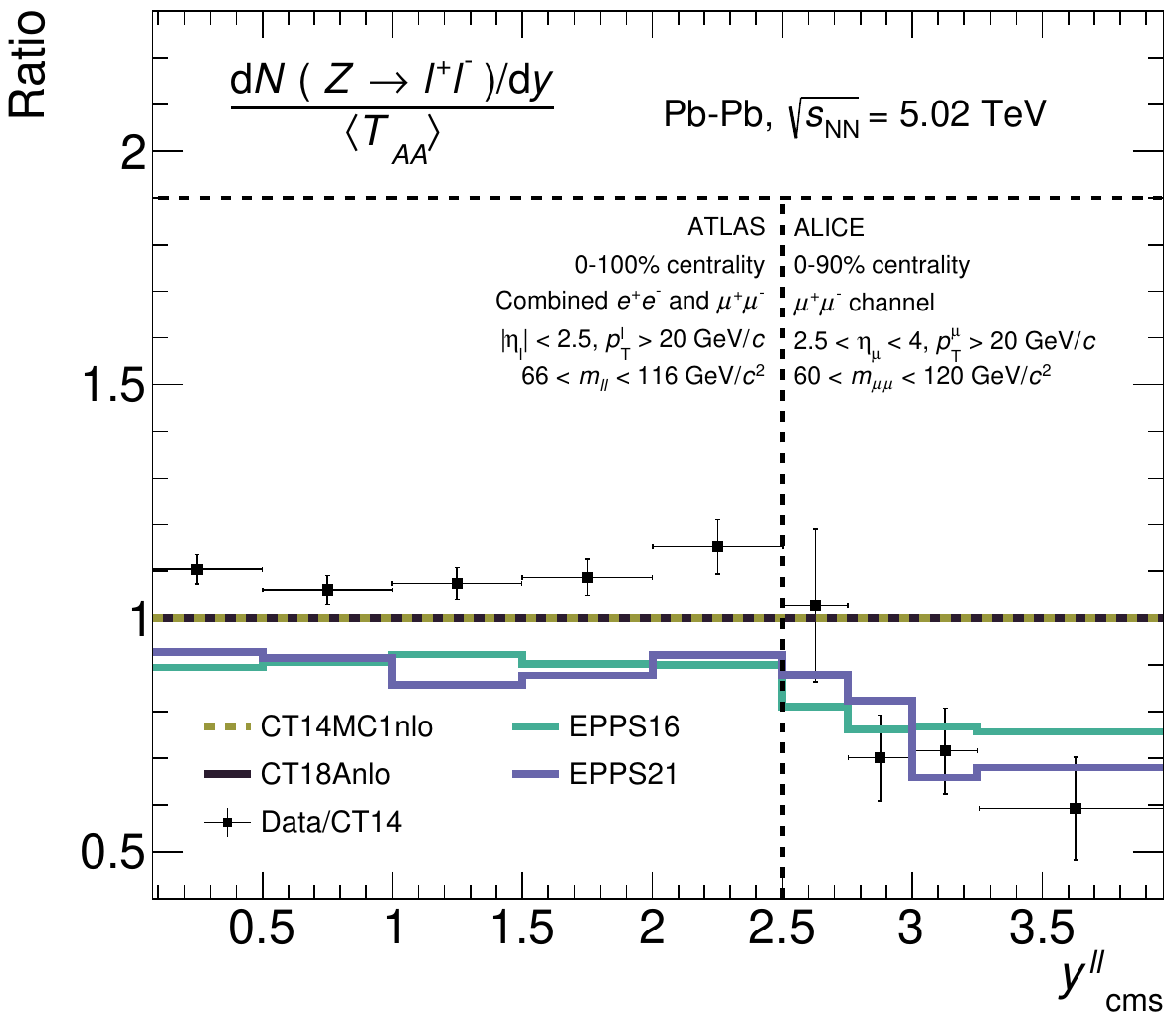} 
  \caption{ The ratio of $Z$-boson yield (black points) as a function of rapidity measured by ALICE and ATLAS in Pb-Pb collisions at $\sqrt{s}=5.02$ TeV to free-nucleon PDF (CT14). The ratios are compared to predictions using different free proton PDFs (CT14NLO, CT18ANLO) and nuclear PDFs (EPPS16, EPPS21).}
  \label{fig:ZBoson}
\end{figure}

\begin{figure*}[ht!]
    \centering
    \includegraphics[scale = 0.8]{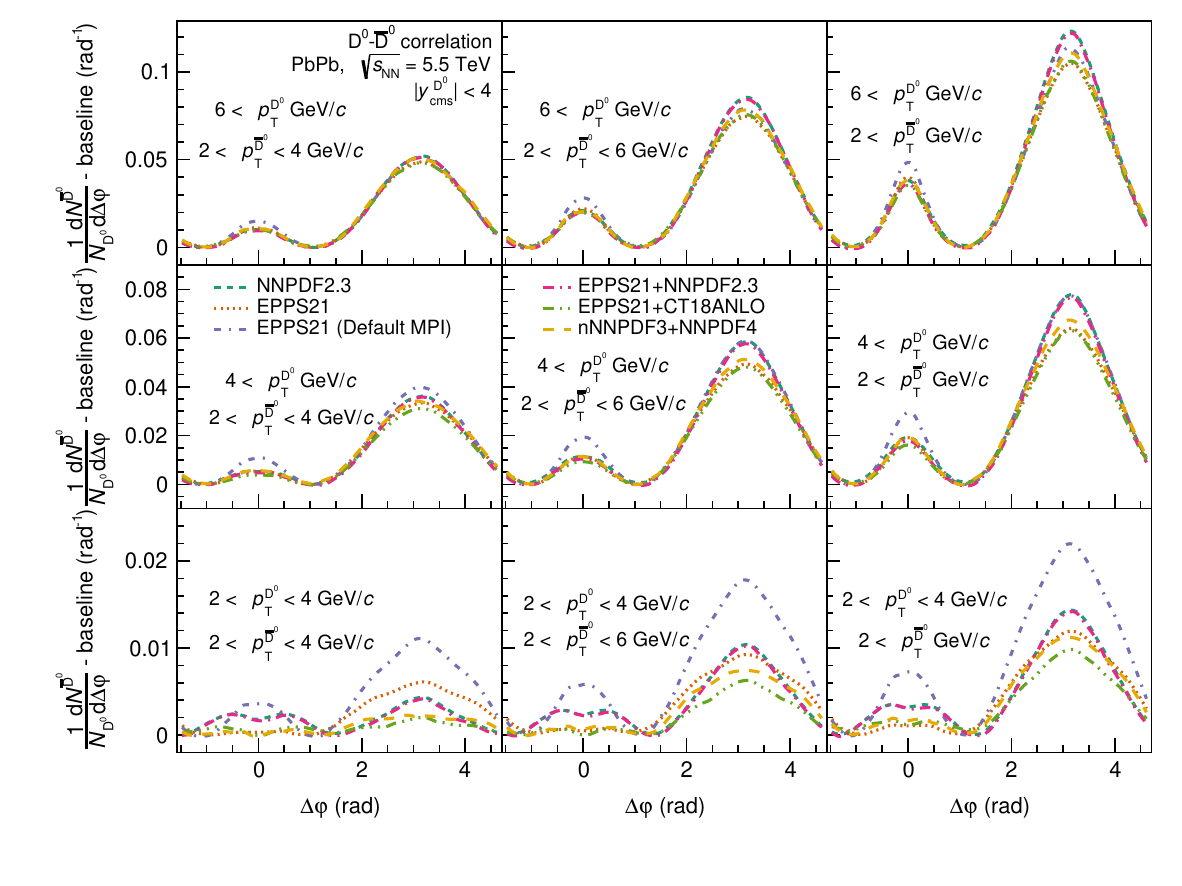}
    \caption{Azimuthal angular correlations of prompt \ddbar mesons from the same hard scattering process in Pb-Pb collisions at $\sqrt{s} = 5.5$ TeV, simulated using PYTHIA8+Angantyr with different PDF and nPDF settings. Results are shown for various \pttD and \ptaD ranges.}
    \label{fig:ddbar_corr}
\end{figure*}

\subsection{Methodology}


In addition to the default proton PDF in PYTHIA8, we append LHAPDF6 to access the recent free proton PDFs (CT18ANLO~\cite{Hou:2019efy}, NNPDF4~\cite{NNPDF:2021njg}) and nuclear PDFs (EPPS21~\cite{Eskola:2021nhw}, nNNPDF3~\cite{AbdulKhalek:2022fyi}). To establish the effect of nPDFs, we examine the production of Z-boson as a function of rapidity. In FIG~\ref{fig:ZBoson}, the ratio between these two PDF settings is compared with the Z-boson production data from the ATLAS and ALICE experiments. To produce Z-bosons, weak processes were enabled using \texttt{WeakSingleBoson:all=ON}, \texttt{WeakDoubleBoson:all=ON}. The rapidity dependence also shows that at large rapidities, nuclear PDFs provide a better description of experimental data than free proton PDF calculations.

\begin{figure*}[ht!]
    \centering
    \includegraphics[scale = 0.8]{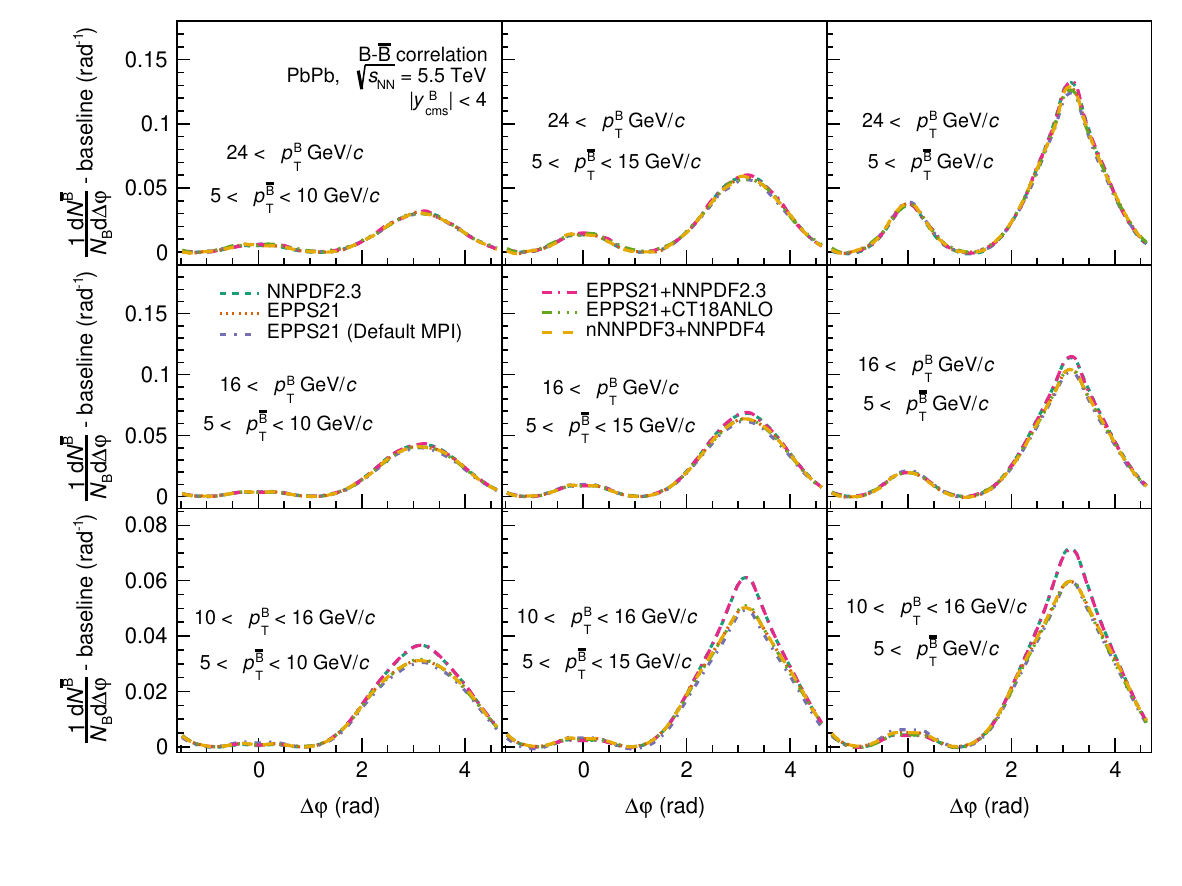}
    \caption{Azimuthal angular correlations of prompt \bbbar mesons from the same hard scattering process in Pb-Pb collisions at $\sqrt{s} = 5.5$ TeV, simulated using PYTHIA8+Angantyr with different PDF and nPDF settings. Results are shown for various \pttB and \ptaB ranges.}
    \label{fig:corr_B}
\end{figure*}

\begin{figure*}[ht!]
    \centering
    \includegraphics[scale = 0.8]{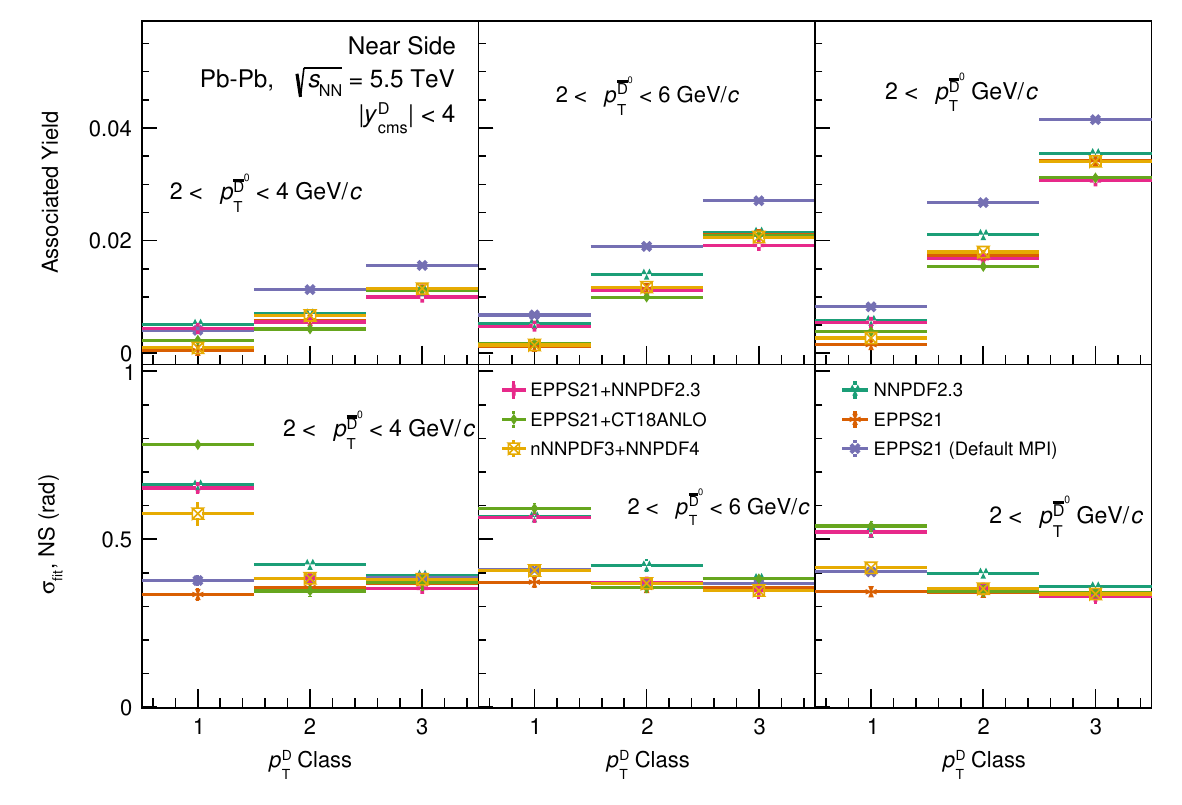}
    \caption{Comparison of near-side yield (top row) and width (bottom row) of azimuthal angular correlations for prompt \ddbar mesons in Pb-Pb collisions at $\sqrt{s} = 5.5$ TeV, simulated using PYTHIA8+Angantyr. Results are presented for different trigger momentum classes (\pttD) and associated momentum ranges (\ptaD), comparing various PDF and nPDF settings.}
    \label{fig:ddbar_ns}
\end{figure*}

\begin{figure*}[ht!]
    \centering
    \includegraphics[scale = 0.8]{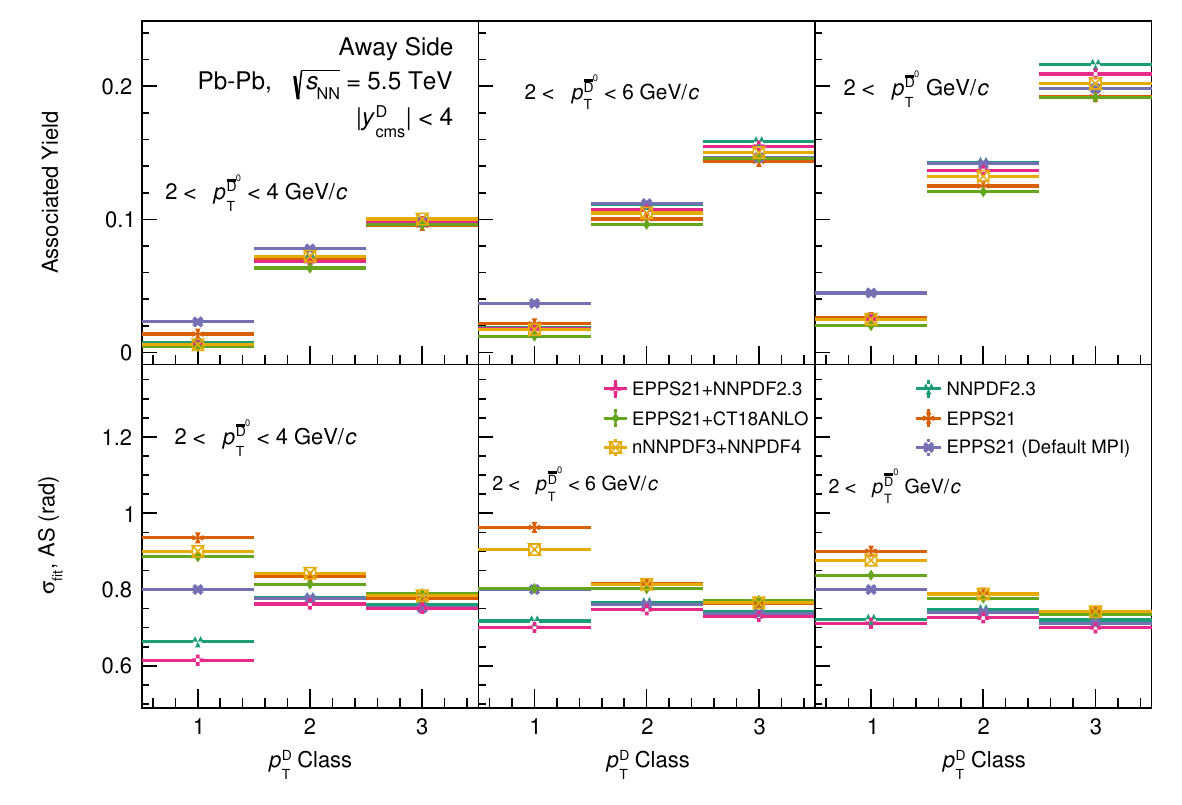}
    \caption{Comparison of away-side yield (top row) and width (bottom row) of azimuthal angular correlations for prompt \ddbar mesons in Pb-Pb collisions at $\sqrt{s} = 5.5$ TeV, simulated using PYTHIA8+Angantyr. Results are presented for different trigger momentum classes (\pttD) and associated momentum ranges (\ptaD), comparing various PDF and nPDF settings.}
    \label{fig:ddbar_as}
\end{figure*}

We simulate Pb-Pb collisions at $\sqrt{s_{NN}} =$ 5.5 TeV for 6 combinations of PDFs. These combinations are NNPDF2.3, EPPS21 (Default MPI), EPPS21, EPPS21+NNPDF2.3, EPPS21+CT18ANLO, and nNNPDF3+NNPDF4. In the first three cases, the mentioned PDFs are used for hard scatterings, ISR, FSR, and MPIs, while in later cases, nPDFs are used for hard scatterings, and free proton PDFs are used for other processes. It has been observed that the choice of PDFs significantly changes the multiplicity distribution of an event. In all cases except for EPPS21 (Default MPI), multiplicity distribution is matched to ALICE midrapidity Pb-Pb data by tuning parameter related to MPIs in PYTHIA8~\cite{Bierlich:2022pfr}. The non-diffractive component of the total cross-section for all soft QCD processes is used with the switch \texttt{SoftQCD: Non-diffractive=on} with MPI, ISR, and FSR to simulate minimum bias events. Open heavy-flavor trigger hadrons are selected by maintaining the ALICE3 acceptance of $\mathrm{|y| < 4.0}$. In the case of prompt \ddbar, the correlation distribution was obtained by correlating $\mathrm{D^0}$ and $\mathrm{\bar{D}^0}$ coming from the same hard scattering. This is ensured by tracing the produced (anti-)charm quark until it hadronizes to $\mathrm{(\bar{D}^{0})D^0}$, through its decay daughter at each stage of fragmentation. To ensure they come from the same scattering, the mothers of $\mathrm{c}$ and $\mathrm{\bar{c}}$ are matched. The same procedure is repeated for $\mathrm{B-\bar{B}}$ meson correlation, but in this case, all B-meson species are used as trigger particles instead of just $\mathrm{B^0}$.

The prompt \ddbar correlation distribution is fitted with two generalized Gaussian functions for the NS peak\footnote[1]{~The two generalized Gaussians are used to estimate the double peak structure forming at the NS.}, a Gaussian function for the AS peak, and a $\mathrm{0^{\text{th}}}$ order polynomial for the baseline identification given as
\begin{align}
f(\Delta\phi) = b + \sum_{i=1}^{2} &\frac{Y_{\rm{NS_i}}\times\beta_{\rm{NS_i}}}{2\alpha_{\rm{NS_i}}\Gamma{(1/\beta_{\rm{NS_i}})}}\times e^{-\left(\frac{\Delta\phi}{\alpha_{\rm{NS_i}}}\right)^{\beta_{\rm{NS_i}}}} \notag \\
& + \frac{Y_{\rm{AS}}}{\sqrt{2\pi}\sigma_{\rm{AS}}}\times e^{-\left(\frac{\Delta\phi - \pi}{\sqrt{2}\sigma_{\rm{AS}}}\right)^{2}}
\label{eq:fit_d}
\end{align}
Here, $Y_{\rm{NS}}$ and $Y_{\rm{AS}}$ are the yields for NS and AS peaks, $\beta_{\rm{NS}}$ is the shape parameter for the NS peak, and $\alpha_{\rm{NS}}$ is related to the $\sigma_{\rm{NS}}$ (width) of the peak as
\begin{equation}
\sigma_{\rm{NS}} = \alpha_{\rm{NS}}\sqrt{\Gamma(3/\beta_{\rm{NS}})/\Gamma(1/\beta_{\rm{NS}})} 
\label{eq:sigma_d}
\end{equation}

In the case of \bbbar, the AS peak exhibits a Gaussian-like structure for low associated \pt ranges, while a triangular nature for high associated \pt ranges inspired us to develop a novel function for the AS peak. The \bbbar \delphi distribution is fitted with two generalized Gaussian functions for the NS peak, the convolution of a Gaussian function and a triangular function for the AS peak, and a $\mathrm{0^{\text{th}}}$ order polynomial for the baseline identification given by

\begin{align}
    f(\Delta\phi) &= b + \sum_{i=1}^{2} \frac{Y_{\rm{NS_i}} \times \beta_{\rm{NS_i}}}{2\alpha_{\rm{NS_i}}\Gamma(1/\beta_{\rm{NS_i}})} \times e^{-\left(\frac{\Delta\phi}{\alpha_{\rm{NS_i}}}\right)^{\beta_{\rm{NS_i}}}} \notag \\
    &\quad + (a - |\Delta\phi - \pi|) \frac{Y_{\rm{AS}}}{\sqrt{2\pi}\sigma_{\rm{AS}}} \times e^{-\left(\frac{\Delta\phi - \pi}{\sqrt{2}\sigma_{\rm{AS}}}\right)^{2}}
    \label{eq:fit_B}
\end{align}

Here, the AS mean is fixed at $\rm{\Delta\phi=\pi}$, and the parameter $``a$'' is responsible for the shape of the distribution. For small values of $``a$'', the AS function behaves like a Gaussian, while for large values of $``a$'', it becomes triangular. The width of the AS distribution is extracted from its standard deviation.

The choice of fitting function for the three heavy flavors is based on the reduced-$\chi^{2}$ of the fit that best describes the shape of the obtained \delphi correlations.


\begin{figure*}[ht!]
    \centering
    \includegraphics[scale = 0.8]{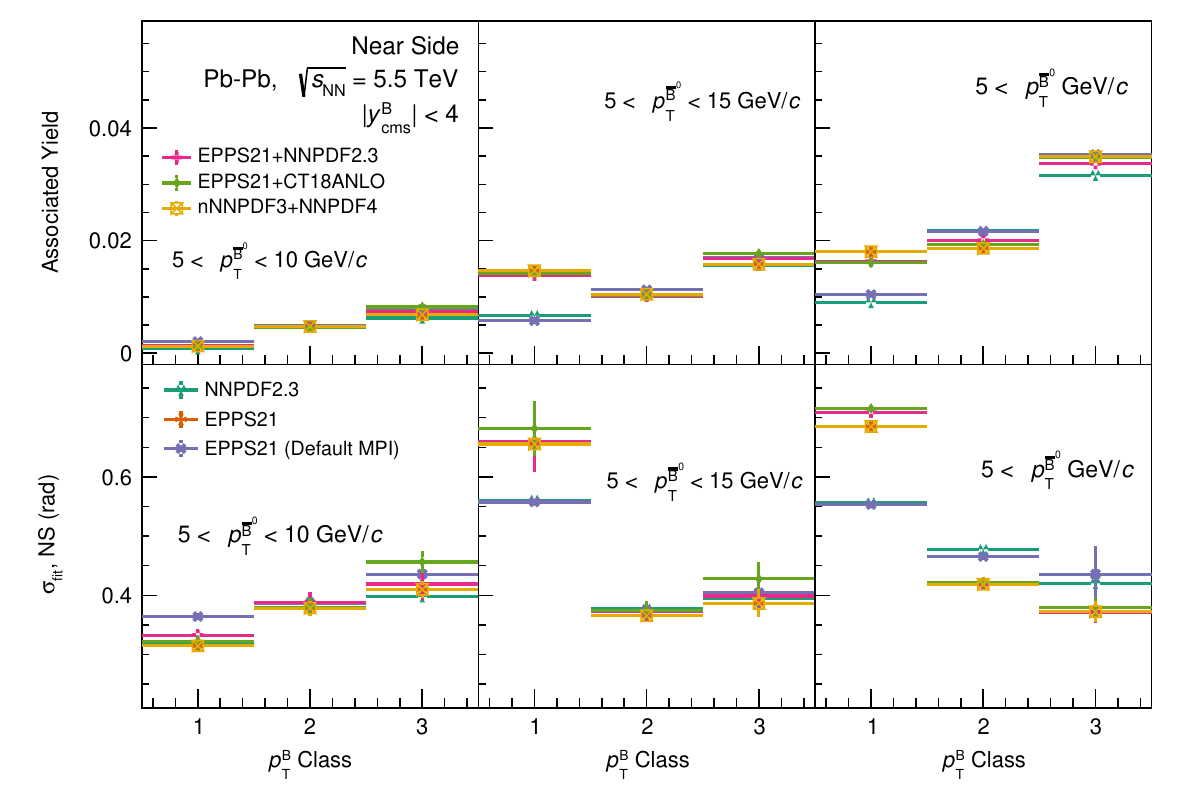}
    \caption{Comparison of near-side yield (top row) and width (bottom row) of azimuthal angular correlations for prompt \ddbar mesons in Pb-Pb collisions at $\sqrt{s} = 5.5$ TeV, simulated using PYTHIA8+Angantyr. Results are presented for different trigger momentum classes (\pttB) and associated momentum ranges (\ptaB), comparing various PDF and nPDF settings.}
    \label{fig:yield_B_ns}
\end{figure*}

\begin{figure*}[ht!]
    \centering
    \includegraphics[scale = 0.8]{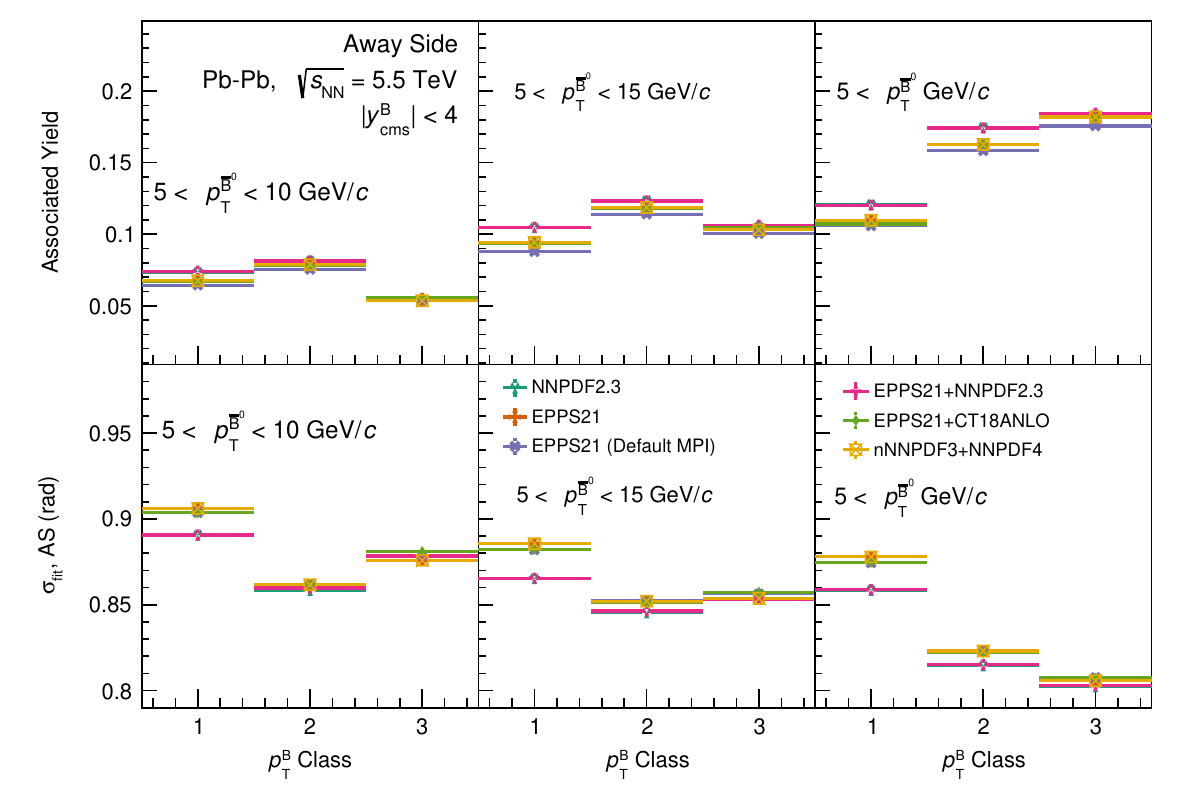}
    \caption{Comparison of away-side yield (top row) and width (bottom row) of azimuthal angular correlations for prompt \ddbar mesons in Pb-Pb collisions at $\sqrt{s} = 5.5$ TeV, simulated using PYTHIA8+Angantyr. Results are presented for different trigger momentum classes (\pttB) and associated momentum ranges (\ptaB), comparing various PDF and nPDF settings.}
    \label{fig:yield_B_as}
\end{figure*}

\begin{figure*}[ht!]
    \centering
    \includegraphics[scale = 0.7]{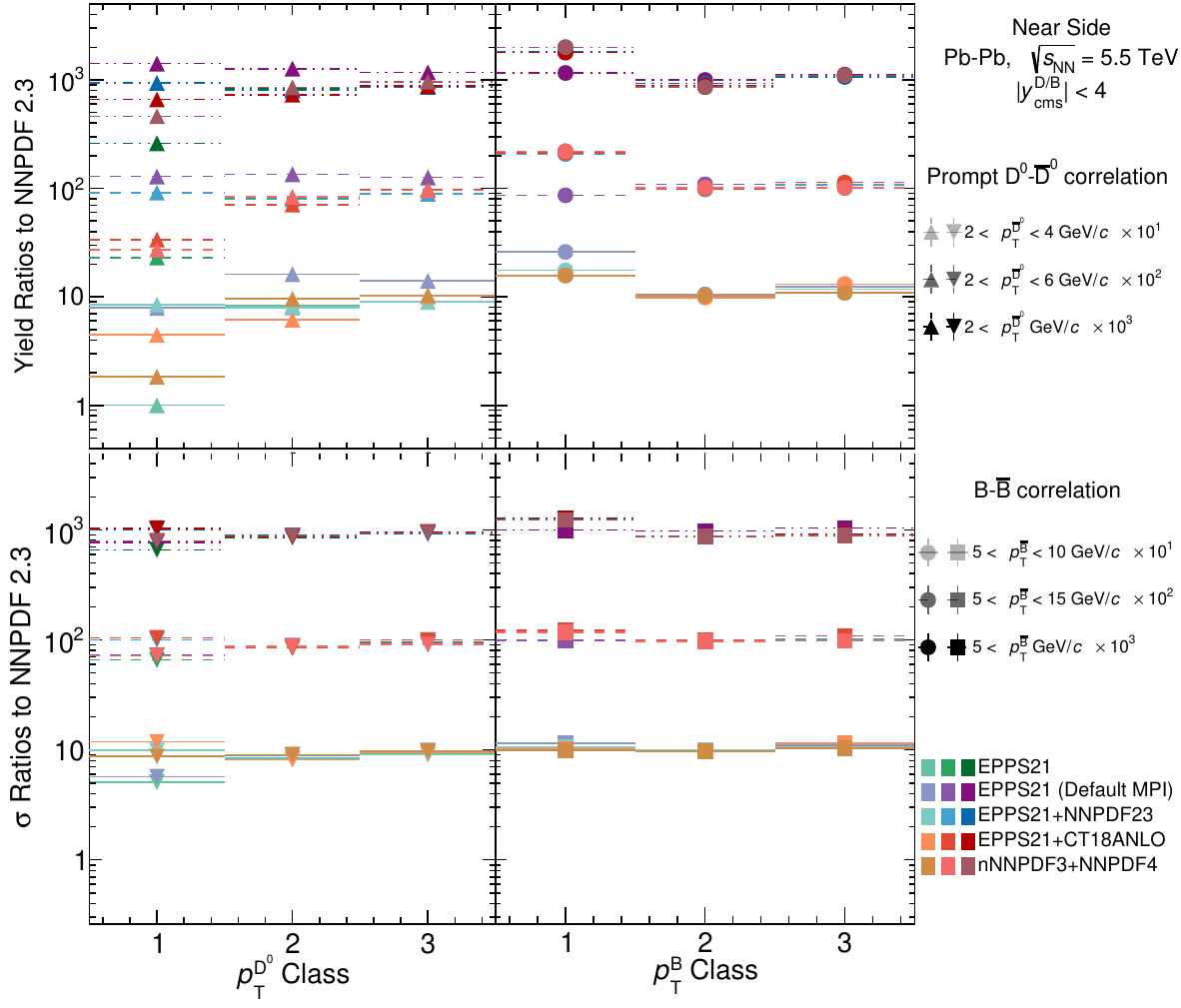}
    \caption{Ratios of near-side yield (top) and width (bottom) of azimuthal angular correlations to the NNPDF2.3 baseline for prompt \ddbar, and \bbbar meson pairs in Pb--Pb collisions at $\sqrt{s_{NN}} = 5.5$ TeV, simulated using PYTHIA8+Angantyr for various PDF and nPDF settings. The left and right columns correspond to prompt $\rm{D^0}$, and B-mesons, respectively. Different line styles represent \pta ranges, scaled for better visualization.}
    \label{fig:ratio_ns}
\end{figure*}

\begin{figure*}[ht!]
    \centering
    \includegraphics[scale = 0.75]{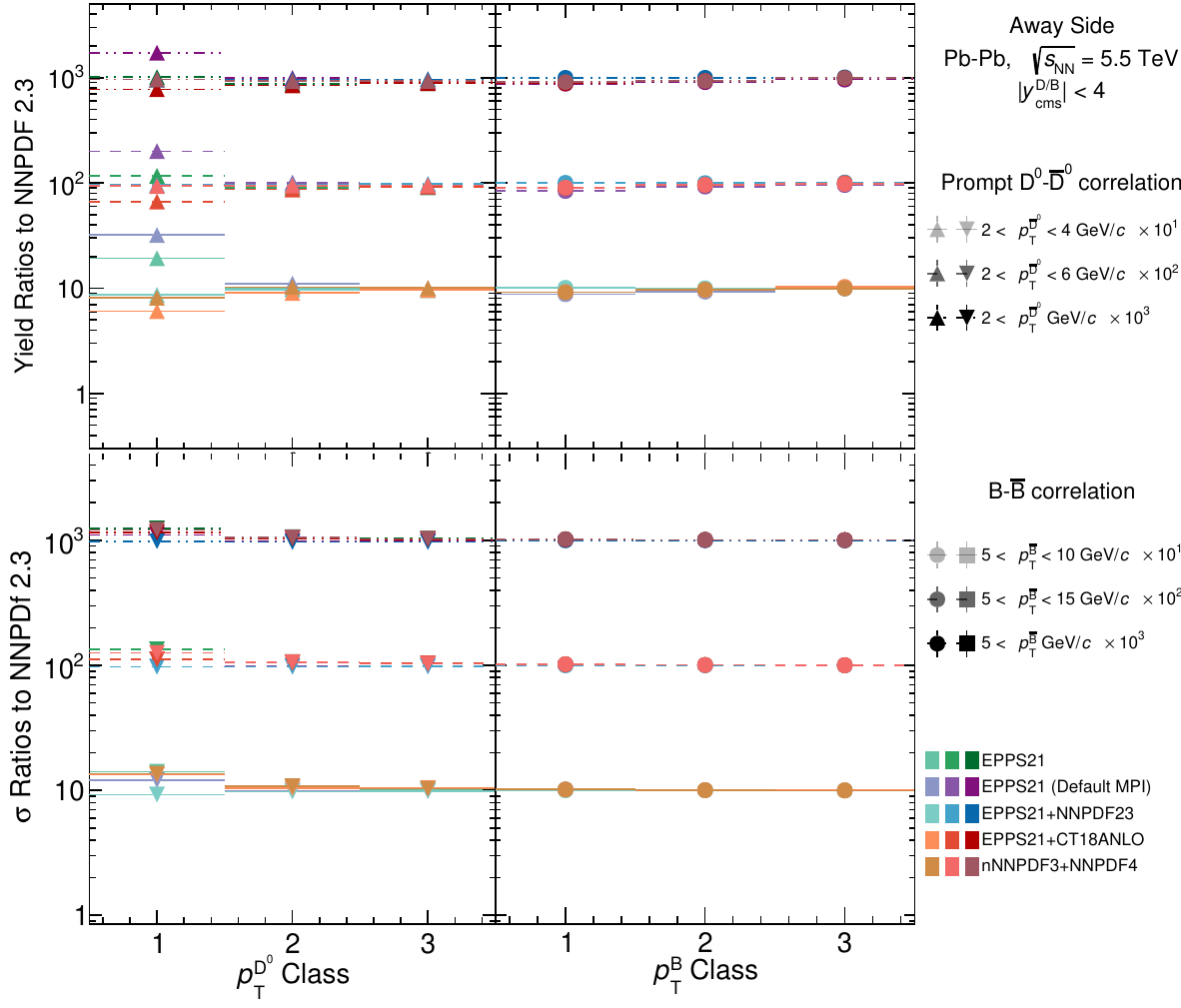}
    \caption{Ratios of away-side yield (top) and width (bottom) of azimuthal angular correlations to the NNPDF2.3 baseline for prompt \ddbar, and \bbbar meson pairs in Pb--Pb collisions at $\sqrt{s_{NN}} = 5.5$ TeV, simulated using PYTHIA8+Angantyr for various PDF and nPDF settings. The left and right columns correspond to prompt $\rm{D^0}$ and B-mesons, respectively. Different line styles represent \pta ranges, scaled for better visualization.}
    \label{fig:ratio_as}
\end{figure*}

\section{Results} 
\label{section4}

The azimuthal angular correlation of \ddbar\ pairs for different trigger and associated transverse momentum (\pt) ranges is presented in FIG.~\ref{fig:ddbar_corr}. Based on the \pt, \dz's are categorized into three classes: Class I (\ptt\ = 2--4 GeV/\textit{c}), Class II (\ptt\ $\geq$ 4 GeV/\textit{c}), and Class III (\ptt\ $\geq$ 6 GeV/\textit{c}). The associated particles are selected within the \pta\ ranges of 2--4, 2--6, and 2--50 GeV/\textit{c}. For B-mesons (FIG.~\ref{fig:corr_B}), the AS distribution is found to dominate over the NS distribution, exhibiting a broader width at the lower trigger and associated \pt\ ranges. Therefore, to investigate the characteristics of azimuthal angular correlations of B mesons arising from different production processes, the trigger particle classes are defined with \ptt\ ranges of 10--16, $\geq$ 16, and $\geq$ 24 GeV/\textit{c} and the associated particles are selected within $|\Delta \eta| < 1$ and \pta\ ranges of 5--10, 10--15, and $> 5$ GeV/\textit{c} which are higher compared to \ddbar.

FIG.~\ref{fig:ddbar_corr} and~\ref{fig:corr_B} show that the NS distribution is smaller than the AS distribution. This is because, at LHC energies, the cross-section for the gluon splitting process, which contributes to the NS distribution, is smaller compared to flavor excitation and pair creation, which predominantly contribute to the AS distribution. 

\begin{table}[ht!]
\begin{tabular}{|c|ccc|}
\hline
\pta (\gev) & \multicolumn{3}{c|}{Mean of NS peaks (rad)}                 \\ \hline
        & \multicolumn{1}{c|}{$\quad\rm{m_1}\quad$}     & \multicolumn{1}{c|}{$\quad\rm{m_2}\quad$}    & $\rm{|m_1 - m_2|}$ \\ \hline
2 < \pt < 4 & \multicolumn{1}{c|}{-0.58} & \multicolumn{1}{c|}{0.56} & 1.14   \\ 
2 < \pt < 6  & \multicolumn{1}{c|}{-0.51} & \multicolumn{1}{c|}{0.46}  & 0.97   \\ 
\pt > 2     & \multicolumn{1}{c|}{-0.43} & \multicolumn{1}{c|}{0.43} & 0.86  \\ \hline
\end{tabular}
\caption{Angular separation between near-side (NS) peaks of \ddbar correlation distribution for different associated \ptaD ranges and \pttD using NNPDF2.3 PDFs.}
\label{tab:table_d_NS}
\end{table}

\begin{figure*}[ht!]
  \centering
  \includegraphics[width=0.8\textwidth]{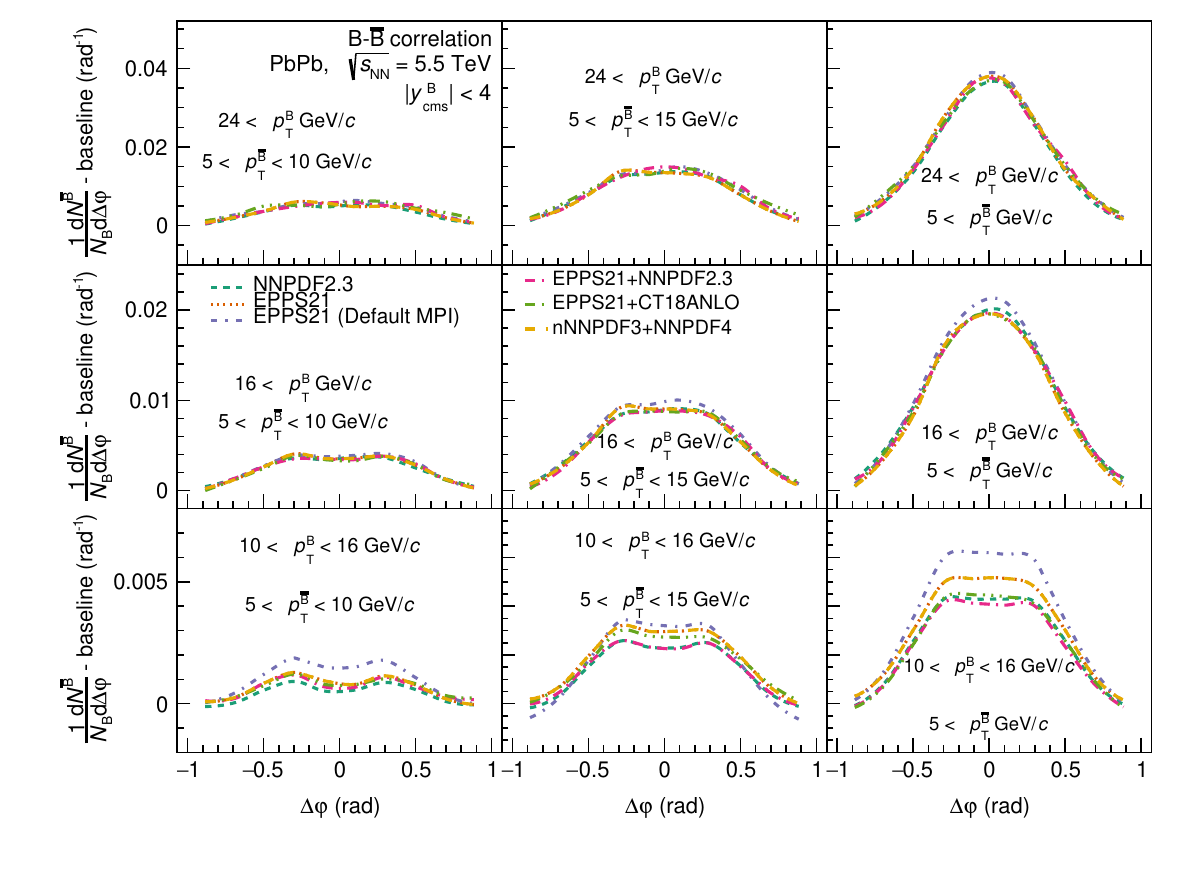} 
  \caption{ Comparison of NS azimuthal angular correlation of \bbbar meson between free proton PDF (NNPDF2.3) and nuclear PDF (EPPS21,nNNPDF3) with PYTHIA8+Angantyr in Pb-Pb collisions at $\rm{\sqrt{s}= 5.5 \, TeV}$ is shown for different  \ptt and \pta ranges. The NS peak is plotted for $\rm{|\Delta \phi|}$ = 1.}
  \label{fig:bbbar_ns}
\end{figure*}

FIG.~\ref{fig:ddbar_corr} further illustrates that, except for EPPS21 (Default MPI), all PDFs give a double-peak structure at NS of correlation distribution in the \ptt range $2<$\ptt$<4$ \gev. This structure arises from $\mathrm{c-\bar{c}}$ production via the gluon splitting process. In FIG.~\ref{fig:corr_B}, a similar double-peak feature is observed for the \bbbar correlation distribution across all \pta classes. FIG.~\ref{fig:bbbar_ns} (Appendix~\ref{sec:appendix}) shows NS of \bbbar correlation distribution for better visualization of double-peak. 

Additionally, it is observed that the angle between the produced $\mathrm{c-\bar{c}}$ ($\mathrm{b-\bar{b}}$) pairs in the gluon splitting process decreases as \pta increases within a given \ptt range, as shown in TABLE~\ref{tab:table_d_NS} and TABLE~\ref{tab:table_b_NS} (Appendix~\ref{sec:appendix}). This behavior can be explained by the fact that, at low \pta, the gluon splitting opening angle is wider, while at higher \pta, the angle narrows due to gluon boost.

\begin{table*}[ht!]
\centering
\renewcommand{\arraystretch}{1.2} 
\setlength{\tabcolsep}{8pt} 

\begin{tabular}{|p{3.5cm}|>{\centering\arraybackslash}p{1.0cm}|
>{\centering\arraybackslash}p{1.0cm}|>{\centering\arraybackslash}p{1.0cm}|
>{\centering\arraybackslash}p{1.0cm}|>{\centering\arraybackslash}p{1.0cm}|
>{\centering\arraybackslash}p{1.0cm}|>{\centering\arraybackslash}p{1.0cm}|
>{\centering\arraybackslash}p{1.0cm}|>{\centering\arraybackslash}p{1.0cm}|}
\hline
\textbf{\ptt (\gev)} & \multicolumn{3}{c|}{\textbf{Class I}} & \multicolumn{3}{c|}{\textbf{Class II}} & \multicolumn{3}{c|}{\textbf{Class III}} \\ \hline 
\textbf{\pta (\gev)} & \textbf{2-4} & \textbf{2-6} & \textbf{$>$2} & \textbf{2-4} & \textbf{2-6} & \textbf{$>$2} & \textbf{2-4} & \textbf{2-6} & \textbf{$>$2} \\ \hline
\multicolumn{10}{|c|}{\textbf{\% change in near side yield with respect to NNPDF2.3}} \\ \hline
\textbf{PDF Settings} & & & & & & & & & \\ \hline
EPPS21+NNPDF2.3 & -15 & -9 & -6 & -21 & -21 & -20 & -10 & -11 & -14 \\ \hline
EPPS21+CT18ANLO &-55 & -66 & -34 & -38 & -29 &-27 & 1 &-2 &-12 \\ \hline
nNNPDF3.0+NNPDF4.0 &-82 &-73 & -54 & -4 &-16 & -15&3 &-4 &-4 \\ \hline
EPPS21 &-90 &-77&-74&-17 &-18 &-18 &1 &-2.0 &-3 \\ \hline
EPPS21 (Default MPI) &-21 & 29& 41& 62 & 35 & 27&40 &26 &17 \\ \hline
\multicolumn{10}{|c|}{\textbf{\% change in near side width with respect to NNPDF2.3}} \\ \hline
EPPS21+NNPDF2.3 &-1 &0 &0 &-10 &-12 &-11 &-10 &-9 &-8 \\ \hline
EPPS21+CT18ANLO &18 &4 &4 &-19 &-15 &-13 &-5 &0 &-4 \\ \hline
nNNPDF3.0+NNPDF4.0 &-13 &-28 &-20 &-10 &-12 &-20 &-3 &-9 &-6 \\ \hline
EPPS21 &-49 &-34 &-34 &-16 &-15 &-14 &-6 &-7 &-6 \\ \hline
EPPS21 (Default MPI) &-43 &-28 &-23 &-10 &-12 &-11 &-1 &-4 &-6 \\ \hline
\end{tabular}

\caption{Impact of different PDF settings on the near-side (AS) yield and width of \ddbar azimuthal angular correlations in Pb--Pb collisions at $\sqrt{s_{NN}} = 5.5$ TeV. Values indicate the percentage deviation from the NNPDF2.3 baseline, presented for various \ptt classes and \pta ranges.}
\label{tab:pdf_comparison_ns}
\end{table*}

\begin{table*}[ht!]
\centering
\renewcommand{\arraystretch}{1.2} 
\setlength{\tabcolsep}{8pt} 

\begin{tabular}{|p{3.5cm}|>{\centering\arraybackslash}p{1.0cm}|
>{\centering\arraybackslash}p{1.0cm}|>{\centering\arraybackslash}p{1.0cm}|
>{\centering\arraybackslash}p{1.0cm}|>{\centering\arraybackslash}p{1.0cm}|
>{\centering\arraybackslash}p{1.0cm}|>{\centering\arraybackslash}p{1.0cm}|
>{\centering\arraybackslash}p{1.0cm}|>{\centering\arraybackslash}p{1.0cm}|}
\hline
\textbf{\ptt (\gev)} & \multicolumn{3}{c|}{\textbf{Class I}} & \multicolumn{3}{c|}{\textbf{Class II}} & \multicolumn{3}{c|}{\textbf{Class III}} \\ \hline 
\textbf{\pta (\gev)} & \textbf{2-4} & \textbf{2-6} & \textbf{$>$2} & \textbf{2-4} & \textbf{2-6} & \textbf{$>$2} & \textbf{2-4} & \textbf{2-6} & \textbf{$>$2} \\ \hline
\multicolumn{10}{|c|}{\textbf{\% change in away side yield with respect to NNPDF2.3}} \\ \hline
\textbf{PDF Settings} & & & & & & & & & \\ \hline
EPPS21+NNPDF2.3 & -14 &-4 &-4 & -3 & -4 & -4 & -1 & -2 & -3\\ \hline
EPPS21+CT18ANLO & -40 & -34 & -22 & -10 & -13 & -15 & -4 & -8 & -11 \\ \hline
nNNPDF3.0+NNPDF4.0 & -19 & -7 &  4 & 3 & -6 & -7 & 1 & -5 & -7\\ \hline
EPPS21 & 92 & 17 & 1 & -1 & -10 & -12 & -4 & -9 & -11 \\ \hline
EPPS21 (Default MPI) & 220 & 100 & 72 & 11 & 1 & -1 & -2 & -8 & -8 \\ \hline
\multicolumn{10}{|c|}{\textbf{\% change in away side width with respect to NNPDF2.3}} \\ \hline
EPPS21+NNPDF2.3 & -7 & -2 & -1 & -2 & -2 & -3 & -1 & -2 & -3 \\ \hline
EPPS21+CT18ANLO & 34 & 12& 16 & 4& 5 & 4 & 4& 4 & 2\\ \hline
nNNPDF3.0+NNPDF4.0 & 36& 26 & 21 & 8 & 6 & 6 & 3 & 3 & 3 \\ \hline
EPPS21 & 41 & 34 & 25& 7& 6 & 6 & 2 & 3 & 3\\ \hline
EPPS21 (Default MPI) & 21 & 11 & 11 & 0 & -1 & -1 & -1 & -1 & -1 \\ \hline
\end{tabular}

\caption{Impact of different PDF settings on the away-side (AS) yield and width of \ddbar azimuthal angular correlations in Pb--Pb collisions at $\sqrt{s_{NN}} = 5.5$ TeV. Values indicate the percentage deviation from the NNPDF2.3 baseline, presented for various \ptt classes and \pta ranges.}
\label{tab:pdf_comparison_as}
\end{table*}

FIG.~\ref{fig:ddbar_ns} and FIG.~\ref{fig:yield_B_ns} present the yield and width of the NS distribution for \dz and B-meson correlations, respectively, while FIG.~\ref{fig:ddbar_as} and FIG.~\ref{fig:yield_B_as} depict the same for the AS distribution. For \dz mesons, the NS yield increases with both \ptt and \pta across all PDF settings. FIG.~\ref{fig:ddbar_ns} shows that, the NS width decreases as a function of \ptt and \pta except in the case of EPPS21. For \bbbar correlations (FIG.~\ref{fig:yield_B_ns}), the NS width exhibits minimal \pta dependence in \ptt Classes II and III, whereas in \ptt Class I, it increases with \pta. For \pta $> 5$ \gev, the NS width decreases with \ptt due to the greater boost of high \pt trigger and associated particles. However, for $5 <$ \pta $< 10$ \gev, the NS width increases with \ptt. In this \pta range, associated particles undergo significant fragmentation and scattering compared to the trigger particle, reducing the correlation strength and resulting in a broader angular distribution. A similar effect is observed for the AS width (FIG.~\ref{fig:yield_B_as}), where \ptt Class III exhibits a broader distribution than Class II. This effect is more pronounced in the AS than in the NS due to the smaller $\Delta p$ of the heavy quarks produced via the pair creation process compared to those from gluon splitting~\cite{Norrbin:2000zc}. However, with the inclusion of higher-\pt associated particles, the expected trend of decreasing width with \ptt is restored. 

FIG.~\ref{fig:ddbar_as} shows that the AS yield increases with \ptt and \pta for \dz mesons. However, for B-mesons (FIG.~\ref{fig:yield_B_as}), the AS yield rises in \ptt classes I and II but decreases in Class III. The difference in AS yield between classes II and III diminishes as higher-\pt associated $\bar{B}$ mesons are included. For \pta $\rm{> 5}$ \gev, the AS yield in \ptt Class III is higher than that of Class II, restoring the usual \pta dependence. These observations can be attributed to the availability of associated particles for \ptt Class III compared to Class II. Associated particles predominantly exhibit $\Delta p \approx 3~\mathrm{GeV}/c$ for pair creation and $\Delta p \approx 5~\mathrm{GeV}/c$ for gluon splitting~\cite{Norrbin:2000zc}. As a result, \ptt Class III has fewer associated particles than \ptt Class II in \pta range 5-10 \gev. In contrast, the NS distribution does not follow this trend due to the larger $\Delta p$ range for the gluon splitting process. This behavior remains consistent across all PDFs and nPDFs under consideration.

As discussed earlier, the choice of nPDFs for all processes significantly reduces the multiplicity distribution due to a lower number of MPIs compared to the default PDF. EPPS21 with the number of MPIs same as default PYTHIA8 settings, exhibits a double-peak structure in the \ptt range $2<$\ptt$<4$ \gev, which is absent when using EPPS21 with default MPI settings. This suggests that the double-peak structure arises from hard MPIs enhancing \dz production at low \pt, making the double peak more pronounced.

A comparison between EPPS21+NNPDF2.3, EPPS21+CT18ANLO, and nNNPDF3+NNPDF4 where former PDFs are used only for hard scatterings and later PDFs for ISR, FSR, and MPIs~\cite{Bierlich:2022pfr}, shows that proton PDF choice for MPI, ISR, and FSR significantly impacts correlation shape, yield, and width. The NS peak shape remains similar across PDFs, but correlation distribution is heavily suppressed for CT18ANLO and NNPDF4 in \ptt Class I compared to NNPDF2.3. Quantitatively, TABLE~\ref{tab:pdf_comparison_ns} show that the NS yield decreases by 30–65\% (CT18ANLO) and 50–80\% (NNPDF4) depending on \pta in \ptt Class I. For \ptt Class II, NS yield decreases by $\approx$ 25-40\% for CT18ANLO, while for NNPDF4 yields do not change more than 20\%. The CT18ANLO does not significantly alter the NS width of the correlation, but NNPDF4 exhibits a 20–30\% narrower correlation for the lowest \ptt range. The AS peak remains consistently smaller than the default PDF for CT18ANLO and NNPDF4 across all \ptt and \pta ranges (FIG~\ref{fig:ddbar_corr}). For the \ptt Class I, yield decreases up to 40\% for CT18ANLO, while deviation remains below 20\% for NNPDF4. The AS width increases by up to 35\% for NNPDF4 and up to 30\% for CT18ANLO as reported in TABLE~\ref{tab:pdf_comparison_as}.

A comparison between NNPDF2.3 and EPPS21 reveals that NS correlations are significantly suppressed, whereas AS correlations are larger for EPPS21 compared to NNPDF2.3~(FIG.~\ref{fig:ddbar_corr}). As reported in TABLE~\ref{tab:pdf_comparison_ns}, the NS yield is 70-90\% less than the default PDFs for \ptt Class I. The AS yield is twice that of the default PDFs for \ptaa but becomes comparable when including high \pt associated particles. For other \ptt ranges, the difference remains below 10\%. As per the TABLE~\ref{tab:pdf_comparison_as}, NS widths decrease, whereas AS widths increase by 20–50\% for EPPS21 in the lowest \ptt range, depending on \pta. In \ptt Class II, NS widths deviate by more than 10\% at the lowest \pta, whereas AS widths remain largely unchanged. These results highlight significant modifications due to the nuclear environment, particularly at low \ptt.

A comparison between NNPDF2.3 and EPPS21+NNPDF2.3 examines the impact of nPDFs on hard scattering in the correlation distribution. The NS and AS distributions remain similar across all \ptt and \pta ranges, indicating that in PYTHIA8, the choice of nPDFs for hard scattering does not affect the azimuthal angular correlation of \ddbar. This suggests that the azimuthal angular distribution of \ddbar is primarily governed by the PDFs used for ISR, FSR, and MPIs.

The panel with \ptt $> 4$ \gev and $2 <$ \pta $< 4$ \gev is critical for studying charm quark thermalization, as collective effects dominate in this range. In Pb–Pb collisions, medium effects can flatten the \ddbar azimuthal angular correlation. TABLE~\ref{tab:pdf_comparison_ns} and TABLE~\ref{tab:pdf_comparison_as} show that NS and AS peaks have identical widths within PDF uncertainties (less than 15\%), indicating that nPDFs have no significant impact on azimuthal broadening in this \pt range for minimum bias Pb–Pb events.

In contrast to \ddbar correlations, \bbbar correlations do not show noticeable sensitivity to PDF variations. This can be due to higher \pt ranges of trigger and associated particles. These higher \pt values correspond to larger Bjorken-x, where PDFs are more consistent across parameterizations, reducing their impact on \bbbar correlations.

 \section{Summary}
 \label{section5}
 In this study, we investigate the impact of state-of-the-art PDFs on the azimuthal angular correlation of \ddbar, and \bbbar. We simulate Pb-Pb collisions at $\rm{\sqrt{s}= 5.5 \, TeV}$ using the PYTHIA8 event generator and construct the azimuthal angular correlations of the open heavy flavor mesons by correlating the heavy quark and anti-quark pairs from the same hard scattering. By delving into this analysis, we seek to uncover potential nuclear influences on thermalization and the production of heavy quarks, both at low and high \pt. The key findings of our investigation are outlined as follows:
 
 \begin{itemize}

     \item A comparison between NNPDF2.3 and multiplicity-corrected EPPS21 implies that nuclear PDFs significantly modify the azimuthal angular correlation of open heavy-flavor mesons compared to free proton PDFs. These modifications influence both the yield and width of NS and AS correlations across all \pta and \ptt ranges by altering ISR, FSR, and MPIs.
     \item The nuclear PDF effects vary with both trigger and associated \pt. These effects are most pronounced at low \ptt and \pta, where the sampled Bjorken-$x$ values are small, and nuclear modifications such as shadowing play a significant role. At high \pt, the nuclear modifications become less significant as PDFs converge at large Bjorken-$x$.
     \item A double-peak structure is observed at the NS in \ddbar and \bbbar azimuthal correlations, providing a clear signature of partonic heavy quark production via the gluon splitting process. The separation between the peaks decreases with increasing \pta due to a stronger initial boost. 
    \item A comparison between EPPS21 and EPPS21 (Default MPI) reveals that increased hard multi-partonic interactions (MPIs) in nPDFs enhance \dz production at low \pt, making the double-peak structure more prominent. This highlights the role of MPI in shaping the azimuthal angular distribution in high-energy nuclear collisions.

    \item The observed modifications in azimuthal angular correlations arise from the choice of nuclear PDFs rather than collective effects. The  \ptt $> 4$ GeV/c and $2 <$ \pta $< 4$ \gev range is crucial for studying charm quark thermalization. Nuclear PDFs alone do not alter the width of the azimuthal correlation significantly in minimum bias Pb–Pb events for these \pt ranges. For lower \pt ranges, the impact of nPDFs needs to be taken into account.

    \item This study introduces a new fitting function for the azimuthal angular correlation of \bbbar mesons, formulated as a convolution of Gaussian and triangular functions. Unlike previous models based on generalized Gaussian or von Mises distributions, this function more accurately captures the evolution of the AS peak from Gaussian-like at low \pt to triangular-like at high \pt. This improved parametrization provides a better description of beauty quark correlation dynamics and could be useful for future studies.

    \item Unlike \ddbar correlations, nuclear PDFs have a weaker effect on \bbbar azimuthal correlations. This may be due to the higher mass of the beauty quark, which reduces sensitivity to initial-state nuclear modifications, or the higher \pt ranges used in \bbbar studies. This suggests that nuclear PDF effects can be overlooked for beauty quark studies at high \pt but remain important for charm.
    
    \item Differential study conducted in this article shows that kinematics of heavy flavor production can be studied in greater detail using azimuthal angular correlations.
 \end{itemize}
 
A detailed study of the azimuthal angular correlation of open heavy flavors provides valuable insights into heavy flavor dynamics in both small and large systems. The findings of this study, based on minimum bias Pb–Pb events, can be extended to centrality-dependent analyses, which may reveal additional nuclear modification effects and offer deeper insights into the complex dynamics of heavy quark production and evolution.

\appendix 

\section{Near side features of \bbbar correlation}
\label{sec:appendix}

To further investigate the NS structure of \bbbar azimuthal correlations, we analyze the peak separation for different PDFs in Pb-Pb collisions at $\sqrt{s} = 5.5$ TeV using the PYTHIA8+Angantyr framework. FIG.~\ref{fig:bbbar_ns} presents the NS azimuthal angular correlation of \bbbar mesons for various \ptt and \pta ranges, highlighting differences between free proton PDFs (NNPDF2.3) and nuclear PDFs (EPPS21, nNNPDF3). The NS peak is examined within $\rm{|\Delta \phi|} = 1$. Table~\ref{tab:table_b_NS} quantifies the separation between the two NS peaks across different kinematic classes. The observed trend suggests that peak separation decreases as \ptt increases, indicating a more collimated correlation due to the larger boost of associated and trigger particles at higher momenta.

    \begin{table}[]
    \begin{tabular}{|c|ccc|}
    \hline
    \pta (\gev)                    & \multicolumn{3}{c|}{Mean of NS peaks (rad)}                 \\ \hline
    Class I                        & \multicolumn{1}{c|}{$\quad\rm{m_1}\quad$}     & \multicolumn{1}{c|}{\quad$\rm{m_2}$\quad\quad}    & \multicolumn{1}{c|}{$\rm{|m_1 - m_2|}$} \\ \hline
    $\rm{5 <}$ \pt $\rm{< 10}$ & \multicolumn{1}{c|}{-0.33} & \multicolumn{1}{c|}{0.33} & 0.66   \\ 
    $\rm{5 <}$ \pt $\rm{< 15}$  & \multicolumn{1}{c|}{-0.35} & \multicolumn{1}{c|}{0.29}  & 0.64   \\
    $\rm{5 <}$ \pt               & \multicolumn{1}{c|}{-0.40} & \multicolumn{1}{c|}{0.25} & 0.65   \\ \hline
    Class II              & \multicolumn{1}{c|}{}     & \multicolumn{1}{c|}{}    & \\ \hline
    $\rm{5 <}$ \pt $\rm{< 10}$ & \multicolumn{1}{c|}{-0.30} & \multicolumn{1}{c|}{0.30} & 0.60   \\
    $\rm{5 <}$ \pt $\rm{< 15}$  & \multicolumn{1}{c|}{-0.30} & \multicolumn{1}{c|}{0.27}  & 0.57   \\
    $\rm{5 <}$ \pt               & \multicolumn{1}{c|}{-0.30} & \multicolumn{1}{c|}{0.25} & 0.55   \\ \hline
    Class III                         & \multicolumn{1}{c|}{}     & \multicolumn{1}{c|}{}    &  \\ \hline
    $\rm{5 <}$ \pt $\rm{< 10}$ & \multicolumn{1}{c|}{-0.27} & \multicolumn{1}{c|}{0.25} & 0.52   \\
    $\rm{5 <}$ \pt $\rm{< 15}$  & \multicolumn{1}{c|}{-0.20} & \multicolumn{1}{c|}{0.20}  & 0.40 \\
    $\rm{5 <}$ \pt               & \multicolumn{1}{c|}{-0.09} & \multicolumn{1}{c|}{0.15} & 0.24   \\ \hline
    \end{tabular}
    \caption{Angular separation between near-side (NS) peaks of \bbbar correlation distribution for different associated \ptaB ranges and \pttB using NNPDF2.3 PDFs.}
    \label{tab:table_b_NS}
    \end{table}

\section*{Acknowledgement} 
S.K. acknowledges the financial support provided by the Council of Scientific and Industrial Research (CSIR) (File No.09/1022(0084)/2019-EMR-I), New Delhi. S.K. is also grateful to the authors of PYTHIA8. R.S. acknowledges the financial support (DST/INSPIRE Fellowship/2017/IF170675) by the DST-INSPIRE program of the Government of India. The authors also thank Fabio Colamaria for fruitful discussions. This work uses computational facilities supported by the DST-FIST scheme via SERB Grant No. SR/FST/PSI-225/2016, by the Department of Science and Technology (DST), Government of India.


\begin{thebibliography}{50}

\bibitem{STAR:2005gfr}
J.~Adams \textit{et al.} [STAR],
Nucl. Phys. A \textbf{757}, 102-183 (2005)
doi:10.1016/j.nuclphysa.2005.03.085
[arXiv:nucl-ex/0501009 [nucl-ex]].

\bibitem{Braun-Munzinger:2015hba}
P.~Braun-Munzinger, V.~Koch, T.~Sch\"afer and J.~Stachel,
Phys. Rept. \textbf{621}, 76-126 (2016)
doi:10.1016/j.physrep.2015.12.003
[arXiv:1510.00442 [nucl-th]].

\bibitem{ALICE:2022wpn}
 [ALICE],
[arXiv:2211.04384 [nucl-ex]].

\bibitem{ALICE:2016fzo}
J.~Adam \textit{et al.} [ALICE],
Nature Phys. \textbf{13}, 535-539 (2017)
doi:10.1038/nphys4111
[arXiv:1606.07424 [nucl-ex]].

\bibitem{NA50:2000brc}
M.~C.~Abreu \textit{et al.} [NA50],
Phys. Lett. B \textbf{477}, 28-36 (2000)
doi:10.1016/S0370-2693(00)00237-9

\bibitem{Bonino:2023icn}
L.~Bonino, M.~Cacciari and G.~Stagnitto,
JHEP \textbf{06}, 040 (2024)
doi:10.1007/JHEP06(2024)040
[arXiv:2312.12519 [hep-ph]].

\bibitem{Kartvelishvili:1977pi}
V.~G.~Kartvelishvili, A.~K.~Likhoded and V.~A.~Petrov,
Phys. Lett. B \textbf{78}, 615-617 (1978)
doi:10.1016/0370-2693(78)90653-6

\bibitem{Andersson:1983jt}
B.~Andersson, G.~Gustafson and B.~Soderberg,
Z. Phys. C \textbf{20}, 317 (1983)
doi:10.1007/BF01407824

\bibitem{ALICE:2021dhb}
S.~Acharya \textit{et al.} [ALICE],
Phys. Rev. D \textbf{105}, no.1, L011103 (2022)
doi:10.1103/PhysRevD.105.L011103
[arXiv:2105.06335 [nucl-ex]].

\bibitem{Zhao:2023nrz}
J.~Zhao, J.~Aichelin, P.~B.~Gossiaux, A.~Beraudo, S.~Cao, W.~Fan, M.~He, V.~Minissale, T.~Song and I.~Vitev, \textit{et al.}
Phys. Rev. C \textbf{109}, no.5, 054912 (2024)
doi:10.1103/PhysRevC.109.054912
[arXiv:2311.10621 [hep-ph]].

\bibitem{Fries:2008hs}
R.~J.~Fries, V.~Greco and P.~Sorensen,
Ann. Rev. Nucl. Part. Sci. \textbf{58}, 177-205 (2008)
doi:10.1146/annurev.nucl.58.110707.171134
[arXiv:0807.4939 [nucl-th]].

\bibitem{Frixione:1997ma}
S.~Frixione, M.~L.~Mangano, P.~Nason and G.~Ridolfi,
Adv. Ser. Direct. High Energy Phys. \textbf{15}, 609-706 (1998)
doi:10.1142/9789812812667\_0009
[arXiv:hep-ph/9702287 [hep-ph]].

\bibitem{Norrbin:2000zc}
E.~Norrbin and T.~Sjostrand,
Eur. Phys. J. C \textbf{17}, 137-161 (2000)
doi:10.1007/s100520000460
[arXiv:hep-ph/0005110 [hep-ph]].

\bibitem{Braun-Munzinger:2003pwq}
P.~Braun-Munzinger, K.~Redlich and J.~Stachel,
doi:10.1142/9789812795533\_0008
[arXiv:nucl-th/0304013 [nucl-th]].

\bibitem{Alberico:2013bza}
W.~M.~Alberico, A.~Beraudo, A.~De Pace, A.~Molinari, M.~Monteno, M.~Nardi, F.~Prino and M.~Sitta,
Eur. Phys. J. C \textbf{73}, 2481 (2013)
doi:10.1140/epjc/s10052-013-2481-z
[arXiv:1305.7421 [hep-ph]].

\bibitem{Fujii:2013yja}
H.~Fujii and K.~Watanabe,
Nucl. Phys. A \textbf{920}, 78-93 (2013)
doi:10.1016/j.nuclphysa.2013.10.006
[arXiv:1308.1258 [hep-ph]].

\bibitem{Kniehl:2007erq}
B.~A.~Kniehl, G.~Kramer, I.~Schienbein and H.~Spiesberger,
Phys. Rev. D \textbf{77}, 014011 (2008)
doi:10.1103/PhysRevD.77.014011
[arXiv:0705.4392 [hep-ph]].

\bibitem{Salajegheh:2019ach}
M.~Salajegheh, S.~M.~Moosavi Nejad, H.~Khanpour, B.~A.~Kniehl and M.~Soleymaninia,
Phys. Rev. D \textbf{99}, no.11, 114001 (2019)
doi:10.1103/PhysRevD.99.114001
[arXiv:1904.08718 [hep-ph]].

\bibitem{Epele:2018ewr}
M.~Epele, C.~Garc\'\i{}a Canal and R.~Sassot,
Phys. Lett. B \textbf{790}, 102-107 (2019)
doi:10.1016/j.physletb.2018.11.069
[arXiv:1807.07495 [hep-ph]].

\bibitem{Hwa:1979pn}
R.~C.~Hwa,
Phys. Rev. D \textbf{22}, 1593 (1980)
doi:10.1103/PhysRevD.22.1593

\bibitem{Biro:1994mp}
T.~S.~Biro, P.~Levai and J.~Zimanyi,
Phys. Lett. B \textbf{347}, 6-12 (1995)
doi:10.1016/0370-2693(95)00029-K

\bibitem{EuropeanMuon:1992pyr}
J.~Ashman \textit{et al.} [European Muon],
Z. Phys. C \textbf{57}, 211-218 (1993)
doi:10.1007/BF01565050

\bibitem{Buckley:2014ana}
A.~Buckley, J.~Ferrando, S.~Lloyd, K.~Nordstr\"om, B.~Page, M.~R\"ufenacht, M.~Sch\"onherr and G.~Watt,
Eur. Phys. J. C \textbf{75}, 132 (2015)
doi:10.1140/epjc/s10052-015-3318-8
[arXiv:1412.7420 [hep-ph]].

\bibitem{LHCb:2017yua}
R.~Aaij \textit{et al.} [LHCb],
JHEP \textbf{10}, 090 (2017)
doi:10.1007/JHEP10(2017)090
[arXiv:1707.02750 [hep-ex]].

\bibitem{LHCb:2022dmh}
R.~Aaij \textit{et al.} [LHCb],
Phys. Rev. Lett. \textbf{131}, no.10, 102301 (2023)
doi:10.1103/PhysRevLett.131.102301
[arXiv:2205.03936 [nucl-ex]].

\bibitem{Kusina:2017gkz}
A.~Kusina, J.~P.~Lansberg, I.~Schienbein and H.~S.~Shao,
Phys. Rev. Lett. \textbf{121}, no.5, 052004 (2018)
doi:10.1103/PhysRevLett.121.052004
[arXiv:1712.07024 [hep-ph]].

\bibitem{ALICE:2020mso}
 [ALICE],
CERN-LHCC-2020-009.

\bibitem{ATLAS:2019maq}
G.~Aad \textit{et al.} [ATLAS],
Phys. Lett. B \textbf{802}, 135262 (2020)
doi:10.1016/j.physletb.2020.135262
[arXiv:1910.13396 [nucl-ex]].

\bibitem{ALICE:2020jff}
S.~Acharya \textit{et al.} [ALICE],
JHEP \textbf{09}, 076 (2020)
doi:10.1007/JHEP09(2020)076
[arXiv:2005.11126 [nucl-ex]].

\bibitem{CMS:2017qjw}
A.~M.~Sirunyan \textit{et al.} [CMS],
Phys. Lett. B \textbf{782}, 474-496 (2018)
doi:10.1016/j.physletb.2018.05.074
[arXiv:1708.04962 [nucl-ex]].

\bibitem{ALICE:2015zhm}
J.~Adam \textit{et al.} [ALICE],
Phys. Lett. B \textbf{754}, 81-93 (2016)
doi:10.1016/j.physletb.2015.12.067
[arXiv:1509.07491 [nucl-ex]].

\bibitem{ALICE:2019hno}
S.~Acharya \textit{et al.} [ALICE],
Phys. Rev. C \textbf{101}, no.4, 044907 (2020)
doi:10.1103/PhysRevC.101.044907
[arXiv:1910.07678 [nucl-ex]].

\bibitem{ALICE:2015vxz}
J.~Adam \textit{et al.} [ALICE],
JHEP \textbf{03}, 081 (2016)
doi:10.1007/JHEP03(2016)081
[arXiv:1509.06888 [nucl-ex]].

\bibitem{ALICE:2018lyv}
S.~Acharya \textit{et al.} [ALICE],
JHEP \textbf{10}, 174 (2018)
doi:10.1007/JHEP10(2018)174
[arXiv:1804.09083 [nucl-ex]].

\bibitem{STAR:2014wif}
L.~Adamczyk \textit{et al.} [STAR],
Phys. Rev. Lett. \textbf{113}, no.14, 142301 (2014)
[erratum: Phys. Rev. Lett. \textbf{121}, no.22, 229901 (2018)]
doi:10.1103/PhysRevLett.113.142301
[arXiv:1404.6185 [nucl-ex]].

\bibitem{ALICE:2023gjj}
S.~Acharya \textit{et al.} [ALICE],
Eur. Phys. J. C \textbf{83}, no.12, 1123 (2023)
doi:10.1140/epjc/s10052-023-12259-3
[arXiv:2307.14084 [nucl-ex]].

\bibitem{STAR:2017kkh}
L.~Adamczyk \textit{et al.} [STAR],
Phys. Rev. Lett. \textbf{118}, no.21, 212301 (2017)
doi:10.1103/PhysRevLett.118.212301
[arXiv:1701.06060 [nucl-ex]].

\bibitem{ALICE:2014qvj}
B.~B.~Abelev \textit{et al.} [ALICE],
Phys. Rev. C \textbf{90}, no.3, 034904 (2014)
doi:10.1103/PhysRevC.90.034904
[arXiv:1405.2001 [nucl-ex]].

\bibitem{ALICE:2013olq}
B.~Abelev \textit{et al.} [ALICE],
Phys. Rev. Lett. \textbf{111}, 102301 (2013)
doi:10.1103/PhysRevLett.111.102301
[arXiv:1305.2707 [nucl-ex]].

\bibitem{ALICE:2020hdw}
S.~Acharya \textit{et al.} [ALICE],
Phys. Rev. Lett. \textbf{126}, no.16, 162001 (2021)
doi:10.1103/PhysRevLett.126.162001
[arXiv:2005.11130 [nucl-ex]].

\bibitem{Beraudo:2014boa}
A.~Beraudo, A.~De Pace, M.~Monteno, M.~Nardi and F.~Prino,
Eur. Phys. J. C \textbf{75}, no.3, 121 (2015)
doi:10.1140/epjc/s10052-015-3336-6
[arXiv:1410.6082 [hep-ph]].

\bibitem{PHENIX:2018wex}
A.~Adare \textit{et al.} [PHENIX],
Phys. Rev. C \textbf{99}, no.5, 054903 (2019)
doi:10.1103/PhysRevC.99.054903
[arXiv:1803.01749 [hep-ex]].

\bibitem{Zhang:2019bkf}
L.~Y.~Zhang, J.~H.~Chen, Z.~W.~Lin, Y.~G.~Ma and S.~Zhang,
Phys. Rev. C \textbf{99}, no.5, 054904 (2019)
doi:10.1103/PhysRevC.99.054904
[arXiv:1904.08603 [nucl-th]].

\bibitem{Zhang:2018ucx}
L.~Y.~Zhang, J.~H.~Chen, Z.~W.~Lin, Y.~G.~Ma and S.~Zhang,
Phys. Rev. C \textbf{98}, no.3, 034912 (2018)
doi:10.1103/PhysRevC.98.034912
[arXiv:1808.10641 [nucl-th]].

\bibitem{ALICE:2019oyn}
S.~Acharya \textit{et al.} [ALICE],
Eur. Phys. J. C \textbf{80}, no.10, 979 (2020)
doi:10.1140/epjc/s10052-020-8118-0
[arXiv:1910.14403 [nucl-ex]].

\bibitem{ALICE:2021kpy}
S.~Acharya \textit{et al.} [ALICE],
Eur. Phys. J. C \textbf{82}, no.4, 335 (2022)
doi:10.1140/epjc/s10052-022-10267-3
[arXiv:2110.10043 [nucl-ex]].

\bibitem{ALICE:2016clc}
J.~Adam \textit{et al.} [ALICE],
Eur. Phys. J. C \textbf{77}, no.4, 245 (2017)
doi:10.1140/epjc/s10052-017-4779-8
[arXiv:1605.06963 [nucl-ex]].

\bibitem{ALICE:2022wwr}
 [ALICE],
[arXiv:2211.02491 [physics.ins-det]].

\bibitem{Sjostrand:2006za}
T.~Sjostrand, S.~Mrenna and P.~Z.~Skands,
JHEP \textbf{05}, 026 (2006)
doi:10.1088/1126-6708/2006/05/026
[arXiv:hep-ph/0603175 [hep-ph]].

\bibitem{Bierlich:2022pfr}
C.~Bierlich, S.~Chakraborty, N.~Desai, L.~Gellersen, I.~Helenius, P.~Ilten, L.~L\"onnblad, S.~Mrenna, S.~Prestel and C.~T.~Preuss, \textit{et al.}
SciPost Phys. Codeb. \textbf{2022}, 8 (2022)
doi:10.21468/SciPostPhysCodeb.8
[arXiv:2203.11601 [hep-ph]].

\bibitem{Bierlich:2018xfw}
C.~Bierlich, G.~Gustafson, L.~L\"onnblad and H.~Shah,
JHEP \textbf{10}, 134 (2018)
doi:10.1007/JHEP10(2018)134
[arXiv:1806.10820 [hep-ph]].

\bibitem{Ball:2016spl}
R.~D.~Ball, E.~R.~Nocera and J.~Rojo,
Eur. Phys. J. C \textbf{76}, no.7, 383 (2016)
doi:10.1140/epjc/s10052-016-4240-4
[arXiv:1604.00024 [hep-ph]].

\bibitem{Regge:1959mz}
T.~Regge,
Nuovo Cim. \textbf{14}, 951 (1959)
doi:10.1007/BF02728177

\bibitem{Brodsky:1973kr}
S.~J.~Brodsky and G.~R.~Farrar,
Phys. Rev. Lett. \textbf{31}, 1153-1156 (1973)
doi:10.1103/PhysRevLett.31.1153

\bibitem{Gribov:1972ri}
V.~N.~Gribov and L.~N.~Lipatov,
Sov. J. Nucl. Phys. \textbf{15}, 438-450 (1972)
IPTI-381-71.

\bibitem{Lipatov:1974qm}
L.~N.~Lipatov,
Yad. Fiz. \textbf{20}, 181-198 (1974)

\bibitem{Altarelli:1977zs}
G.~Altarelli and G.~Parisi,
Nucl. Phys. B \textbf{126}, 298-318 (1977)
doi:10.1016/0550-3213(77)90384-4

\bibitem{Dokshitzer:1977sg}
Y.~L.~Dokshitzer,
Sov. Phys. JETP \textbf{46}, 641-653 (1977)

\bibitem{Barshay:1975zz}
S.~Barshay, C.~B.~Dover and J.~P.~Vary,
Phys. Rev. C \textbf{11}, 360-369 (1975)
doi:10.1103/PhysRevC.11.360

\bibitem{Ethier:2020way}
J.~J.~Ethier and E.~R.~Nocera,
Ann. Rev. Nucl. Part. Sci. \textbf{70}, 43-76 (2020)
doi:10.1146/annurev-nucl-011720-042725
[arXiv:2001.07722 [hep-ph]].

\bibitem{NNPDF:2017mvq}
R.~D.~Ball \textit{et al.} [NNPDF],
Eur. Phys. J. C \textbf{77}, no.10, 663 (2017)
doi:10.1140/epjc/s10052-017-5199-5
[arXiv:1706.00428 [hep-ph]].

\bibitem{Ball:2012cx}
R.~D.~Ball, V.~Bertone, S.~Carrazza, C.~S.~Deans, L.~Del Debbio, S.~Forte, A.~Guffanti, N.~P.~Hartland, J.~I.~Latorre and J.~Rojo, \textit{et al.}
Nucl. Phys. B \textbf{867}, 244-289 (2013)
doi:10.1016/j.nuclphysb.2012.10.003
[arXiv:1207.1303 [hep-ph]].

\bibitem{Hou:2019efy}
T.~J.~Hou, J.~Gao, T.~J.~Hobbs, K.~Xie, S.~Dulat, M.~Guzzi, J.~Huston, P.~Nadolsky, J.~Pumplin and C.~Schmidt, \textit{et al.}
Phys. Rev. D \textbf{103}, no.1, 014013 (2021)
doi:10.1103/PhysRevD.103.014013
[arXiv:1912.10053 [hep-ph]].

\bibitem{Cacciari:1998it}
M.~Cacciari, M.~Greco and P.~Nason,
JHEP \textbf{05}, 007 (1998)
doi:10.1088/1126-6708/1998/05/007
[arXiv:hep-ph/9803400 [hep-ph]].

\bibitem{Dulat:2015mca}
S.~Dulat, T.~J.~Hou, J.~Gao, M.~Guzzi, J.~Huston, P.~Nadolsky, J.~Pumplin, C.~Schmidt, D.~Stump and C.~P.~Yuan,
Phys. Rev. D \textbf{93}, no.3, 033006 (2016)
doi:10.1103/PhysRevD.93.033006
[arXiv:1506.07443 [hep-ph]].

\bibitem{Aivazis:1993pi}
M.~A.~G.~Aivazis, J.~C.~Collins, F.~I.~Olness and W.~K.~Tung,
Phys. Rev. D \textbf{50}, 3102-3118 (1994)
doi:10.1103/PhysRevD.50.3102
[arXiv:hep-ph/9312319 [hep-ph]].

\bibitem{Eskola:2021nhw}
K.~J.~Eskola, P.~Paakkinen, H.~Paukkunen and C.~A.~Salgado,
Eur. Phys. J. C \textbf{82}, no.5, 413 (2022)
doi:10.1140/epjc/s10052-022-10359-0
[arXiv:2112.12462 [hep-ph]].

\bibitem{AbdulKhalek:2022fyi}
R.~Abdul Khalek, R.~Gauld, T.~Giani, E.~R.~Nocera, T.~R.~Rabemananjara and J.~Rojo,
Eur. Phys. J. C \textbf{82}, no.6, 507 (2022)
doi:10.1140/epjc/s10052-022-10417-7
[arXiv:2201.12363 [hep-ph]].

\bibitem{NNPDF:2021njg}
R.~D.~Ball \textit{et al.} [NNPDF],
Eur. Phys. J. C \textbf{82}, no.5, 428 (2022)
doi:10.1140/epjc/s10052-022-10328-7
[arXiv:2109.02653 [hep-ph]].

\bibitem{NNPDF:2023tyk}
R.~D.~Ball \textit{et al.} [NNPDF],
Phys. Rev. D \textbf{109}, no.9, L091501 (2024)
doi:10.1103/PhysRevD.109.L091501
[arXiv:2311.00743 [hep-ph]].

\bibitem{Singh:2021edu}
R.~Singh, Y.~Bailung and A.~Roy,
Phys. Rev. C \textbf{105}, no.3, 035202 (2022)
doi:10.1103/PhysRevC.105.035202
[arXiv:2108.08626 [nucl-th]].

\bibitem{Sjostrand:2007gs}
T.~Sjostrand, S.~Mrenna and P.~Z.~Skands,
Comput. Phys. Commun. \textbf{178}, 852-867 (2008)
doi:10.1016/j.cpc.2008.01.036
[arXiv:0710.3820 [hep-ph]].

\bibitem{Skands:2014pea}
P.~Skands, S.~Carrazza and J.~Rojo,
Eur. Phys. J. C \textbf{74}, no.8, 3024 (2014)
doi:10.1140/epjc/s10052-014-3024-y
[arXiv:1404.5630 [hep-ph]].

\bibitem{Campbell:2022qmc}
J.~M.~Campbell, M.~Diefenthaler, T.~J.~Hobbs, S.~H\"oche, J.~Isaacson, F.~Kling, S.~Mrenna, J.~Reuter, S.~Alioli and J.~R.~Andersen, \textit{et al.}
SciPost Phys. \textbf{16}, no.5, 130 (2024)
doi:10.21468/SciPostPhys.16.5.130
[arXiv:2203.11110 [hep-ph]].

\bibitem{Corke:2010yf}
R.~Corke and T.~Sjostrand,
JHEP \textbf{03}, 032 (2011)
doi:10.1007/JHEP03(2011)032
[arXiv:1011.1759 [hep-ph]].

\bibitem{Lonnblad:2021fyl}
L.~L\"onnblad,
Nucl. Phys. A \textbf{1005}, 121873 (2021)
doi:10.1016/j.nuclphysa.2020.121873

\bibitem{Corke:2009tk}
R.~Corke and T.~Sjostrand,
JHEP \textbf{01}, 035 (2010)
doi:10.1007/JHEP01(2010)035
[arXiv:0911.1909 [hep-ph]].

\bibitem{Diehl:2011yj}
M.~Diehl, D.~Ostermeier and A.~Schafer,
JHEP \textbf{03}, 089 (2012)
[erratum: JHEP \textbf{03}, 001 (2016)]
doi:10.1007/JHEP03(2012)089
[arXiv:1111.0910 [hep-ph]].

\bibitem{Sjostrand:2017cdm}
T.~Sj\"ostrand,
Adv. Ser. Direct. High Energy Phys. \textbf{29}, 191-225 (2018)
doi:10.1142/9789813227767\_0010
[arXiv:1706.02166 [hep-ph]].

\bibitem{Andersson:1983ia}
B.~Andersson, G.~Gustafson, G.~Ingelman and T.~Sjostrand,
Phys. Rept. \textbf{97}, 31-145 (1983)
doi:10.1016/0370-1573(83)90080-7

\bibitem{Khoze:1994fu}
V.~A.~Khoze and T.~Sjostrand,
Phys. Lett. B \textbf{328}, 466-476 (1994)
doi:10.1016/0370-2693(94)91506-7
[arXiv:hep-ph/9403394 [hep-ph]].

\bibitem{Gieseke:2012ft}
S.~Gieseke, C.~Rohr and A.~Siodmok,
Eur. Phys. J. C \textbf{72}, 2225 (2012)
doi:10.1140/epjc/s10052-012-2225-5
[arXiv:1206.0041 [hep-ph]].

\bibitem{Rathsman:1998tp}
J.~Rathsman,
Phys. Lett. B \textbf{452}, 364-371 (1999)
doi:10.1016/S0370-2693(99)00291-9
[arXiv:hep-ph/9812423 [hep-ph]].

\bibitem{Bierlich:2015rha}
C.~Bierlich and J.~R.~Christiansen,
Phys. Rev. D \textbf{92}, no.9, 094010 (2015)
doi:10.1103/PhysRevD.92.094010
[arXiv:1507.02091 [hep-ph]].

\bibitem{Christiansen:2015yqa}
J.~R.~Christiansen and P.~Z.~Skands,
JHEP \textbf{08}, 003 (2015)
doi:10.1007/JHEP08(2015)003
[arXiv:1505.01681 [hep-ph]].

\bibitem{Bierlich:2016smv}
C.~Bierlich, G.~Gustafson and L.~L\"onnblad,
JHEP \textbf{10}, 139 (2016)
doi:10.1007/JHEP10(2016)139
[arXiv:1607.04434 [hep-ph]].

\end{thebibliography}

 \end{document}